\documentclass[useAMS,usenatbib]{mn2e}

\usepackage{psfig}
\usepackage{graphicx}
\usepackage{times}
\usepackage{natbib}

\newif\ifAMStwofonts
\AMStwofontstrue

\newcommand{\be}{\begin{equation}}
\newcommand{\ee}{\end{equation}}
\newcommand{\ba}{\begin{eqnarray}}
\newcommand{\ea}{\end{eqnarray}}
\newcommand{\brr}{\begin{array}}
\newcommand{\err}{\end{array}}
\newcommand{\bc}{\begin{center}}
\newcommand{\ec}{\end{center}}

\newcommand{\vel}{\,{\rm km\,s^{-1}}}

\newcommand{\mincir}{\raise
  -2.truept\hbox{\rlap{\hbox{$\sim$}}\raise5.truept \hbox{$<$}\ }}
\newcommand{\magcir}{\raise
  -2.truept\hbox{\rlap{\hbox{$\sim$}}\raise5.truept \hbox{$>$}\ }}
\newcommand{\siml}{\raise
  -2.truept\hbox{\rlap{\hbox{$\sim$}}\raise5.truept \hbox{$<$}\ }}
\newcommand{\simg}{\raise
  -2.truept\hbox{\rlap{\hbox{$\sim$}}\raise5.truept \hbox{$>$}\ }}

\title[Origin of intracluster stars in cosmological simulations] {The
Importance of Mergers for the Origin of Intracluster Stars in
Cosmological Simulations of Galaxy Clusters}

\author[G.~Murante et al.]  {Giuseppe Murante$^{1,2}$, Martina
Giovalli$^{3}$, Ortwin Gerhard$^4$, Magda Arnaboldi$^{1,5}$,\\~\\
\LARGE{\rm Stefano Borgani$^{2,6,7}$, Klaus Dolag$^8$} \\~\\ 
$^1$ INAF,
Osservatorio Astronomico di Torino, Strada Osservatorio 20, I-10025
Pino Torinese (Italy) (murante@to.astro.it)\\ 
$^2$ Dipartimento di
Astronomia dell'Universit\'a di Trieste, Via Tiepolo 11, I-34131
Trieste (Italy) \\ 
$^3$ Dipartimento di Fisica Generale ``Amedeo
Avogadro'', Universit\'adegli Studi di Torino, Torino (Italy)
(giovalli@to.astro.it)\\ 
$^{4}$ Max-Planck-Institut f\"ur
Extraterrestrische Physik, Giessenbachstrasse, Garching bei M\"unchen,
D-85741. Germany (gerhard@exgal.mpe.mpg.de)\\ 
$^5$ ESO -
Karl-Schwarzschild-Str. 2D-85748 Garching bei M\"unchen
(arnaboldi@to.astro.it)\\ 
$^6$ INAF, Osservatorio Astronomico di
Trieste, Via Tiepolo 11, I-34131 Trieste (Italy)
(borgani@ts.astro.it)\\ 
$^7$ INFN, National Institute for Nuclear
Physics, Trieste (Italy)\\ 
$^8$ Max-Planck-Institut f\"ur Astrophysik,
Karl-Schwarzschild-Strasse, Garching bei M\"unchen, D-85741. Germany
 (kdolag@mpa-garching.mpg.de)\\ 
}

\begin{document}

\maketitle

\label{firstpage}

\begin{abstract}
We study the origin of the diffuse stellar component (DSC) in 117
galaxy clusters extracted from a cosmological hydrodynamical
simulation.  We identify all galaxies present in the simulated
clusters at 17 output redshifts, starting with $z=3.5$, and then build
the family trees for all the $z=0$ cluster galaxies.  The most massive
cluster galaxies show complex family trees, resembling the merger
trees of dark matter halos, while the majority of other cluster
galaxies experience only one or two major mergers during their entire
life history.  Then for each diffuse star particle identified at
$z=0$, we look for the galaxy to which it once belonged at an earlier
redshift, thus linking the presence of the diffuse stellar component
to the galaxy formation history.

The main results of our analysis are: (i) On average, half of the DSC
star particles comes from galaxies associated with the family tree of
the most massive galaxy (BCG), one quarter comes from the family trees
of other massive galaxies, and the remaining quarter from dissolved
galaxies. I.e., the formation of the DSC is parallel to the build-up
of the BCG and other massive galaxies. (ii) Most DSC star particles
become unbound during mergers in the formation history of the BGCs and
of other massive galaxies, independent of cluster mass. Our results
suggest that the tidal stripping mechanism is responsible only for a
minor fraction of the DSC. (iii) At cluster radii larger than 250
$h^{-1}$ kpc, the DSC fraction from the BCG is reduced and the largest
contribution comes from the other massive galaxies; in the cluster
outskirts, galaxies of all masses contribute to the DSC. (iv) The DSC
does not have a preferred redshift of formation: however, most DSC
stars are unbound at $z<1$.  (v) The amount of DSC stars at $z=0$ does
not correlate strongly with the global dynamical history of clusters,
and increases weakly with cluster mass.

\end{abstract}


\section{Introduction}

Observations of diffuse intracluster light and individual intracluster
stars in nearby clusters
\citep{Magda02,Magda03,Magda04,Feld03,MihosVirgo,Ortwin} and at
intermediate redshift \citep{Gonzales2000,Feld04,Zibetti} indicate
that a substantial fraction of stars becomes unbound from galaxies as
these fall towards the densest parts of their cluster environment.

The radial distribution of the intracluster light (ICL) in galaxy
clusters is observed to be more centrally concentrated than that of
the cluster galaxies \citep{Zibetti}, a result which was predicted
from cosmological hydrodynamical simulations of galaxy clusters
\citep[][M04 hereafter]{M04}.  \cite{Zibetti} also find that the
surface brightness of the ICL correlates both with BCG luminosity and
with cluster richness, while the fraction of the total light in the
ICL is almost independent of these quantities. Other observations
indicate an increase of the relative fraction of diffuse stars from
the mass scale of loose groups (less than 2 \%, \cite{Castro-Rodr},
\cite{FeldIAU03})
to that of Virgo-like clusters \citep[$\approx 5-10$\%]{Feld03, Magda03,
MihosVirgo} up to the most massive clusters \citep[10-20 \% or
higher]{Gonzales2000,Feld2002,GalYam,Feld04,Krick06}. 

The origin and evolution of this diffuse stellar component (DSC) is
currently unknown and several mechanism are being investigated. The
ICL may be produced by stripping and disruption of galaxies as they
pass through the central regions of relaxed clusters
\citep{ByrdValt,Gned03}. Other mechanisms are the stripping of stars
from galaxies during the initial formation of clusters
\citep{Merritt84}; creation of stellar halos in galaxy groups, that
later fall into massive clusters, and then become unbound
\citep{Mihos04,Rudick}; stripping of stars during high--speed galaxy
encounters in the cluster environment \citep{Moore96}. Evidence for
ongoing stripping from elliptical galaxies in clusters was
presented by \cite{Cypr06}.

In parallel, numerical simulations have been performed to investigate
the properties the DSC in galaxy clusters within the current
cosmological models. \cite{Napo03} used Dark-Matter (DM) only
simulations, and identified the stellar component using the DM
particles as tracers. For the first time, M04 used a $\Lambda$CDM
cosmological hydrodynamical simulation, including radiative cooling
and star formation, to quantify the amount and the distribution of the
DSC in a set of 117 clusters. \cite{Fabio} and \cite{SommerLarsen}
found a DSC in their simulated single clusters. \cite{Fabio}
discussed the origin of the DSC: they found a correlation
between the cluster growth and the increase in the DSC mass, and that
both massive and small galaxies contribute to its formation.

Recently, \cite{Rudick} performed collisionless simulations where
high--resolution model galaxies were inserted in their dark matter halos
at a given redshift, and then their common evolution in a cluster was
followed from that time on. A DSC was formed, and \cite{Rudick} found
that the cluster DSC grows with the accretion of groups during the
cluster history.

In this work, we focus on the formation mechanism of the ICL in a
cosmological hydrodynamical simulation \citep[][M04]{Borg}.
The formation of galaxies and their subsequent dynamical evolution in
a time dependent gravitational potential is a direct consequence of
the hierarchical assembly process of cosmic structures.  Using a large
($192^3 h^{-3}$ Mpc$^3$) volume simulation, we study a statistically
significant ensemble of galaxy clusters and follow how stars become
unbound from galaxies during the evolution of clusters as a function of
cosmic time. We also address the stability of our results against
numerical resolution by carrying out the same analysis on three
clusters from this set, which were re-simulated at a substantially
improved force and mass resolution.

The plan of the paper is as follows: in Section~\ref{simclus} we describe our
numerical simulations and in Section~\ref{skidid} we give details on the
galaxy identification and properties. In Section~\ref{dscsection} we describe
the identification of the diffuse stellar component (DSC).  In
Section~\ref{sdcorigin} we present the link between galaxy histories and the
formation of the DSC; in Section~\ref{reseffect} we discuss how resolution and
other numerical effect may affect our results; in Section~\ref{secunbound} we
discuss the dynamical mechanisms that unbind stars from galaxies in clusters
and compare with the statistical analysis of the cosmological simulation
performed in the previous Sections.  In Section~\ref{concl} we summarise our
results and give our conclusions.

\section{The simulated clusters}
\label{simclus}
The clusters analysed in this paper are extracted from the large
hydrodynamical simulation (LSCS) of a ``concordance'' $\Lambda$CDM
cosmological model ($\Omega_m=0.3$, $\Omega_\Lambda=0.7$, $\Omega_{\rm
b}=0.019\,h^{-2}$, $h=0.7$ and $\sigma_8=0.8$). This simulation is
presented in \cite{Borg} and we refer to that paper for
additional details. The LSCS is carried out with the massively
parallel Tree+SPH code {\small GADGET2} \citep{GADGET, GADGET2}, and
follows $480^3$ dark matter particles and as many gas particles in a
periodic box of size $192\, h^{-1}$ Mpc. Accordingly, the mass
resolution is $m_{\rm dm}=4.6 \times 10^9\, h^{-1} M_\odot$, $m_{\rm
gas}=6.9 \times 10^8 h^{-1} M_\odot$ and $m_{\rm star}=3.465 \times
10^8 h^{-1} M_\odot$. The Plummer--equivalent softening length for the
gravitational force is set to $\epsilon=7.5\, h^{-1}$ kpc, fixed in
physical units from $z=0$ to $z=2$, while being fixed in co-moving
units at higher redshift. The SPH softening length of the gas is
allowed to shrink to half the value of the gravitational force
softening. The simulation includes radiative cooling, the effect of a
photo--ionising uniform UV background, star formation using a
sub-resolution multi-phase model for the interstellar medium (Springel
\& Hernquist 2003), feedback from supernovae (SN) explosions,
including the effect of galactic outflows. The velocity of these
galactic winds is fixed to $v_w\simeq 340\vel$, which corresponds to
50\% efficiency for SN to power the outflows.

Clusters are identified at $z=0$ using a standard friends-of-friends
(FOF) algorithm, with a linking length of $0.15$ times the mean dark
matter inter-particle separation. We identify 117 clusters in the
simulation with $M_{FOF}>10^{14}h^{-1}M_\odot$. Cluster centres are
placed at the position of the DM particle having the minimum value of
the gravitational potential. For each cluster, the virial mass
$M_{\rm vir}$ is defined as the mass contained within a radius
encompassing an average density equal to the virial density,
$\rho_{\rm vir}$, predicted by the top--hat spherical collapse
model. For the assumed cosmology, $\rho_{\rm vir}\simeq 100\rho_c$,
where $\rho_c$ is the critical cosmic density \citep[e.g.]{Eke96}.

To test the effects of numerical resolution on the final results, we
select three clusters, having virial masses $M_{\rm vir} =1.6,2.5,2.9
\times 10^{14} h^{-1}$M$_\odot$, and re-simulate them twice with
different resolution. While the first, lower-resolution simulation is
carried out at the same resolution as the parent simulation, the
second simulation had a mass resolution $45 \times$ higher, with a
correspondingly smaller softening parameter, $\epsilon=2.1\, h^{-1}$
kpc. These re-simulations are performed using more efficient SN
feedback, with a wind velocity $v_w\simeq 480\vel$. A detailed
description of these re--simulations is provided by \cite{BorgNum}.

\section{Identifying galaxies in a cluster with SKID} 
\label{skidid}

The identification of substructures inside halos is a longstanding
problem, which is not uniquely solved.  In the present work, we need
to identify galaxies in the simulations from the distribution of star
particles which fill the volume of the cosmological simulation.

In the LSCS, ``galaxies'' are defined as self---bound, locally
over-dense structures, following the procedure in M04, which is based
on the publicly available SKID algorithm\footnote{See {\tt
http://www-hpcc.astro.washington.edu/tools/skid.html} } \citep{SK}.  At
a given redshift, once the star particles have been grouped by SKID,
we classify as galaxies only those groups which contain at least 32
bound star particles. There is a degree of uncertainty in the galaxy
identification by SKID, as in other similar identification algorithms,
which comes in from the assignment of those star particles which are
located in its outskirts of each self-bound object.  The main
advantage of this identification algorithm is that it provides a
dynamically--based, automated, operational way to decide whether a
star particle belongs to a gravitationally bound object or not.
Additional details of the galaxy identification algorithm and on our
tests are given in the Appendix.

We expect that, once a self-bound structure of luminous particles has
been formed at a given redshift, most of its mass will remain in bound
structures, for all subsequent redshifts. However, it may happen that
a group of particles classified as a ``galaxy'' at one output
redshift, with a number of particles just above the specified minimum
particle threshold for structure identification, may fall below this
limit at the next redshift output. This may occur, for example,
because the group is evaporated by interaction with the environment.
Following \cite{Springel2001MT}, we call structures that can be
identified only at one output redshift {\sl volatile}, and do not
consider them further.

All star particles that never belong to any galaxies identified in the
selected redshift outputs are also assigned to this volatile
class. Such star particles either do not belong to any bound structure
already at the first output redshift, at $z=3.5$, or they form in a
galaxy {\sl and} become unbound between two simulation output
redshifts. In both cases, since we cannot assign those stars to the
history of any galaxies, we cannot determine their dynamical origin.

An important issue in our study concerns the reliability of the
simulated galaxy population.  If galaxies are under-dense, they can
easily lose stars or be completely disrupted as a consequence of
numerical effects. In simulations, low-mass galaxies may have typical
sizes of the order of the adopted softening parameter, so that their
internal mass density is underestimated. Therefore at the low-mass
end, we expect that our simulated galaxies will have an internal
density which is an {\it increasing} function of galaxy mass. On the
other hand, numerical effects should be less important for the more
massive galaxies.

\begin{figure}
\psfig{file=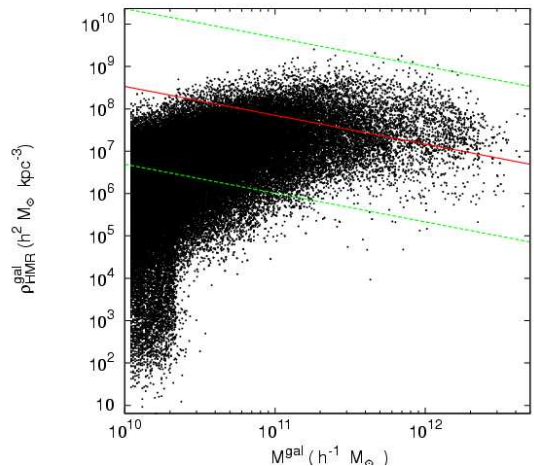,height=7.cm,angle= -90}
\caption{ The mean galaxy stellar density inside the half-mass-radius
  as a function of galaxy mass. All galaxies identified in the parent
  simulation at all 17 redshift outputs are shown. The solid line
  shows an estimate of the observed galaxy densities from SDSS data
  (see text). The dotted lines show the densities corresponding to the
  $3 \sigma$ scatter reported in \citet{Shen}.  }
\label{galdens}
\end{figure}

To investigate this issue, we evaluate the stellar density of all the
simulated galaxies at the half-mass-radius, and plot these in
Figure~\ref{galdens} as a function of the galaxy mass, combining all
redshift outputs.
For real galaxies, the internal density of (early--type) galaxies is a
decreasing function of their mass, as shown most recently by
\cite{Shen}, who measured the size distribution for 140,000 galaxies
from the Sloan Digital Sky Survey. We use their measured size
distribution to estimate the observed galaxy densities within the
half-mass-radius. For this purpose, we take the expressions for a
Hernquist profile in \citep{Hern} to relate Sersic half-light radii to
three-dimensional half-mass radii, and then convert the Sersic
size-stellar mass relation for early-type galaxies of \cite{Shen} to a
relation between stellar mass and mean density within the half-mass
radius.  The solid line in Fig.~\ref{galdens} represents the resulting
estimate of the mean galaxy density, with the dotted lines limiting
the $3 \sigma$ scatter of the size distribution as reported in
\cite{Shen}.  The dots in Fig.~\ref{galdens} show the equivalent mean
densities of our simulated cluster galaxies.  In what follows, we use
the lower $3 \sigma$ envelope to estimate the minimum acceptable
galaxy densities $\rho_{\rm min}(M_{\rm gal})$.  Galaxies with density
lower than this minimum density are discarded and also classified as
volatile.  From Fig.~\ref{galdens} we note that the observed trend of
decreasing density with increasing mass is recovered in our
(lower-resolution) parent simulation for galaxy masses $\magcir
10^{11} h^{-1} M_\odot$.

In order to quantify the effect of volatile galaxies on our final
results, we tested other density thresholds, namely (i) one
corresponding to $1 \sigma$ scatter in the $R_e$ distribution, (ii) a
fixed value of $\rho = 5 \cdot 10^6 h^2 M_\odot$/kpc$^3$, as well as
(iii) a galaxy mass threshold, $M = 6 \cdot 10^{10} h^{-1}
M_\odot$. Our results remain qualitatively unchanged when either of
these criteria is adopted.

\section{Identifying the Diffuse Stellar Component} 
\label{dscsection}

The star formation model implemented in our simulations is based on a
gas--density threshold criterion \citep{SpringHern03}. This
ensures that stars can only form inside existing gravitational
potential wells, so that star formation does not take place outside DM
halos. Thus DSC stars must have become unbound from their parent
galaxies sometime after their formation. Therefore in our analysis, we
define as diffuse stellar component (DSC) all those star particles
which (i) do not belong to any self-bound galaxy at $z=0$, (ii) were
part of a non-volatile structure at earlier redshifts whose
density exceeded the minimum density for its mass as defined above.

In surface brightness measurements of the DSC, sometimes a
distinction is attempted between the component associated with the
halo of the central dominant (cD) galaxy and the intra--cluster light,
which fills the whole cluster region. Quoting from \cite{Uson91}: 
''{\it whether this diffuse light is called cD envelope or
diffuse intergalactic light is a matter of semantics: it is a diffuse
component distributed with elliptical symmetry at the canter of the
cluster potential}''. In our analysis, we will not make such a
distinction: all star particles that do not belong to any self-bound
galaxy at $z=0$, including the cD galaxy identified by SKID, are part
of the diffuse stellar component if they were once part of a
non-volatile, above minimum-density structure.

The part of the DSC contributed by galaxies which have a central
density lower than $\rho_{\rm min}$ is not considered in our analysis,
because it is most likely affected by numerical effects. These
low-density structures include a population of extremely low--density
objects found by SKID at the very low mass end, many of them
representing a mis--identification of SKID due to their small number
of particles ($<100$). However, by discarding the contribution from
low-density and volatile galaxies, we may also neglect a possibly
genuine contribution to the DSC from a population of low--mass
galaxies. Because of this, our estimate of the diffuse light fraction
in the simulated clusters may be an underestimate, although we believe
the corresponding bias to be relatively small; we shall discuss this
issue in Sect.~\ref{reseffect}.

\begin{figure}
\psfig{file=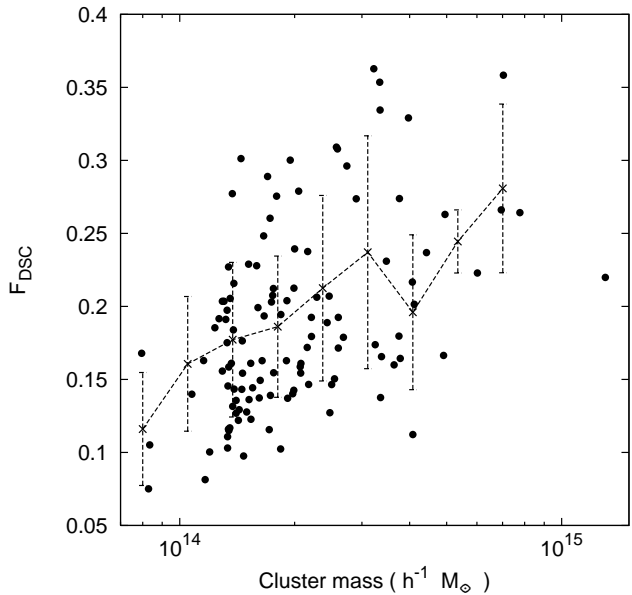,height=8.cm,angle= -90}
\caption{The fraction of stellar mass in the DSC relative to the total
  stellar mass as a function of the cluster virial mass. Dots are for
  the 117 clusters in our parent simulation. The crosses show the
  average values of this ratio in different mass bins, with the
  error bars indicating the r.m.s. scatter within each bin.  }
\label{newsdcfrac}
\end{figure}

Figure~\ref{newsdcfrac} shows the fraction $F_{\rm DSC} = M^*_{\rm
DSC}/M^*_{\rm tot}$, where $M^*_{\rm DSC}$ is the stellar mass
in the DSC and $M^*_{\rm tot}$ is the total stellar mass found within
$R_{\rm vir}$ for each cluster in the parent simulation, as a function
of the cluster mass. In this computation, the diffuse star particles
from volatile galaxies have been discarded. We report on the
fractions of DSC stars in all the steps of our selection procedure in
Table~\ref{clustab}.

Consistent with the results shown in M04, we find: 1) The fraction of
DSC relative to the total stellar light in clusters 
increases with cluster mass, albeit with a large scatter, and 2)
the DSC fraction in the simulated clusters is in the range
$0.1 < F_{\rm DSC}^{obs} < 0.4$. Both results are broadly consistent with
the observed trends and values \citep{MagdaIAU, Aguerri2005, Aguerri2006}; 
however, a direct comparison between observed and measured
values of $F_{\rm DSC}$ is only qualitative, because simulations provide
the volume--averaged mass fraction directly, while this is not true
with the observed DSC fractions.

\section{Tracing the origin of the DSC}
\label{sdcorigin}

The large set of simulated clusters extracted from the cosmological
simulation allows us to perform a statistical study of the origin of
the DSC. From Fig.~\ref{newsdcfrac}, it is clear that clusters with
similar mass can have rather different amounts of DSC at $z=0$. For
this reason, we will first address the general trends in the origin of
the DSC that are independent of the characteristics and dynamical
history of individual clusters, such as the redshift at which the most
of the DSC becomes unbound, and from which galaxies the intracluster stars
mainly originate. We will then investigate whether significant
differences in the production of the DSC can be found between clusters
belonging to different mass classes, and discuss the robustness of our
results against numerical resolution.

We study the origin of the DSC by adopting the following strategy: we
follow back in time all the particles in the DSC component at $z=0$
within each cluster's virial radius and associate them with bound
structures present at any earlier redshifts.  For all clusters and the
17 redshift outputs (from $z=0$ to $z=3.5$), we compile the list of
all galaxies as described in Sect.~\ref{skidid}. Subsequently, for
each DSC particle at $z=0$, we check whether it belong to any of these
galaxies at earlier redshift. If no galaxy is found, the DSC particle
is discarded, because we cannot establish its origin.
If a galaxy is found, then there are three options:
\begin{itemize}  
\item This galaxy has a central density larger than the adopted
  threshold and it belongs to the ``family tree'' of a galaxy
  identified at $z=0$ (see the next subsection); the DSC particle is
  then associated with that family tree.
\item This galaxy has a central density larger than the adopted
  threshold, but it does not belong to a family tree of any galaxy at
  $z=0$; the DSC particle is then considered to come from a
  ``dissolved'' galaxy.
\item This galaxy has a central density below the adopted
  threshold and is thus considered as ``volatile'';  the DSC particle
  is then discarded.
\end{itemize}
In this way, the progenitors of all retained DSC particles can be
found.

\subsection{Building the family trees of galaxies}

We build the merger trees of all galaxies identified at $z=0$, and
refer to them as ``family trees'' to distinguish them from the standard DM
halo merger trees.  

The ``family trees'' are built as follows. For each output redshift $z_{i+1}$
of the simulations, we follow all the DM, star and gas particles within the
virial radius of the identified cluster at $z=0$.  We build catalogs of all
galaxies from the corresponding star and DM particles distributions. For a
given galaxy identified at redshift $z_{i+1}$, we tag all its star particles
and track them back to the previous output redshift $z_i$. We then make a list
of the subset of all identified galaxies at $z_i$ which contain the tagged
particles belonging to the specified galaxy at $z_{i+1}$.

We define a galaxy $G_i$, at output redshift $z_i$, to be progenitor of
a galaxy $G_{j}$ at the next output redshift $z_{i+1}$ if it
contains at least a fraction $g$ of all the stars ending up in
$G_{j}$. The definition of progenitor depends on the fraction $g$.
Our tests show that the number of galaxies identified as progenitors
is stable for $g$ values varying in the range 0.3--0.7. The value
adopted for our analysis is $g=0.5$, which is the same value adopted
in several reconstructions of the DM halo merger trees presented in
the literature \citep{KauffProc, Springel2001MT, Wechsler2002}.

\begin{figure*}
\centerline{
\includegraphics[viewport=0 0 166 298, width=0.33\linewidth, height=8cm]
		{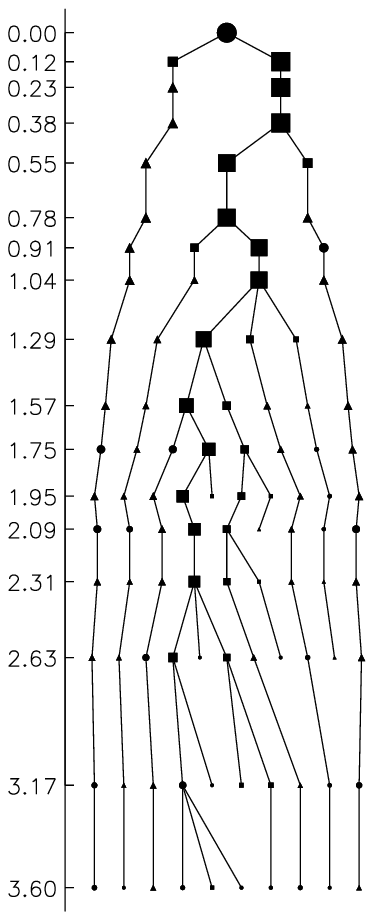}
\includegraphics[viewport=0 0 166 298, width=0.33\linewidth, height=8cm]
		{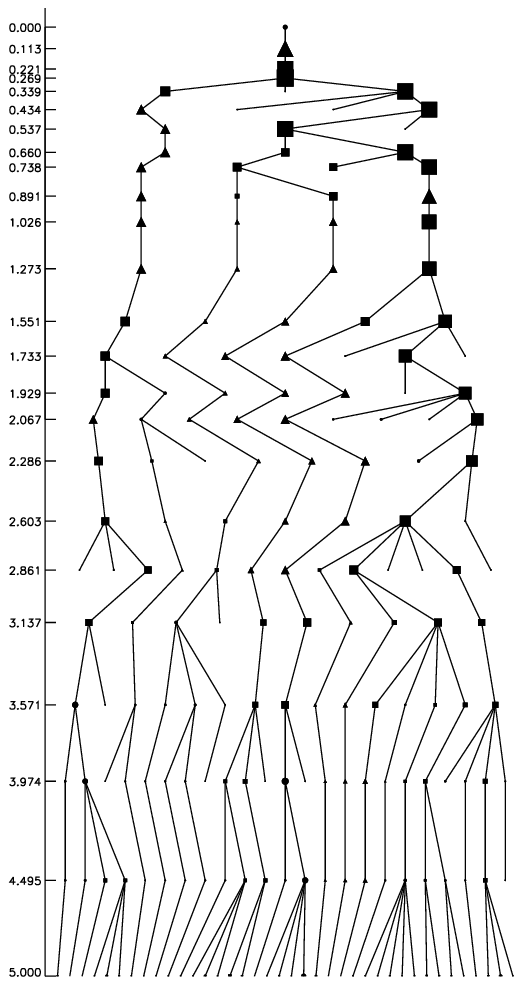}
\includegraphics[viewport=15 0 90 298, width=0.15\linewidth, height=8cm]
		{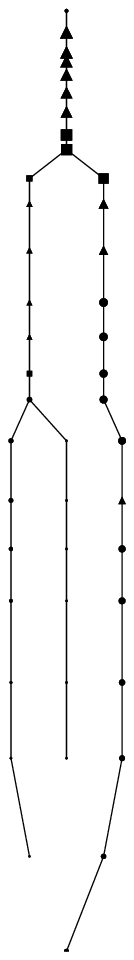}
\includegraphics[viewport=15 0 90 298, width=0.15\linewidth, height=8cm]
		{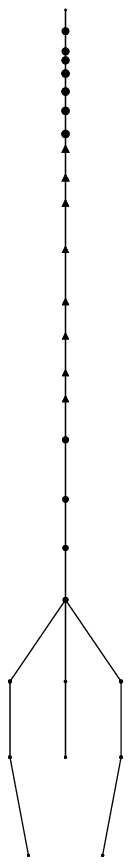}
}
\caption{Left: Family tree of the cD galaxy of cluster A in the
  low-resolution simulation (see Table \ref{clustab}).  Right: Family
  trees of the cD galaxy, the third-most massive galaxy, and a
  lower-mass galaxy in the high-resolution re-simulation of the same
  cluster. -- The size of symbols is proportional to the logarithm of
  the mass of the galaxies at the corresponding redshift.  Shown on
  the vertical axis on the left are the output redshifts used to
  reconstruct the family trees; these are different in both
  simulations. A galaxy in these trees is considered a progenitor of
  another galaxy if at least 50\% of its stars are bound to its
  daughter galaxy, according to the SKID algorithm.  Many more
  galaxies can be identified in the high-resolution simulation at
  similar redshift. The cD family tree is characterised by one
  dominant branch with a number of other branches merging into it, at
  both resolutions.  Squares and triangles represent our
  classification of ``merging'' and ``stripping'' events, see
  Section~\ref{DSCmerging}. Circles correspond to redshift at which the
  galaxy is not releasing stars to the DSC. }
\label{famtreelow}
\end{figure*}

We then build the family trees for all galaxies found at $z=0$ in all
the 117 clusters of our sample. Given the adopted mass threshold of 32
star particle per galaxy, this amounts to an overall number of 1816
galaxies at redshift $z=0$, and 71648 galaxies in all redshift outputs.  

Figure~\ref{famtreelow} shows the family tree of the cD galaxy of a
cluster having virial mass $M{\rm vir}=1.6 \times 10^{14} h^{-1}
M_\odot$ (cluster A in Table~\ref{clustab}).  The cD galaxy family
tree is complex and resembles a typical DM halo merger tree, with the
cD being the result of a number of mergers between pre--existing
galaxies. Other galaxies have a much simpler formation history, with
fewer or no mergers of luminous objects.  This is
illustrated in the right part of Fig.~\ref{famtreelow} which shows
merger trees from the high-resolution re-simulation of the same
cluster, for the cD galaxy, the third-most massive galaxy in the
cluster, and a low-mass galaxy.  In more massive clusters, galaxies
whose family trees are intermediate between that of the cD and the
third-most massive galaxy can also be found. They are however
among the most massive galaxies in their cluster, and they are often
the most massive galaxy of an infalling subcluster, which has not
merged completely with the main cluster yet.

Once the family trees of all galaxies in our clusters at $z=0$ are
built, we then analyse the formation history of the DSC.

\subsection {The epoch of formation of the DSC}

\begin{figure}
\psfig{file=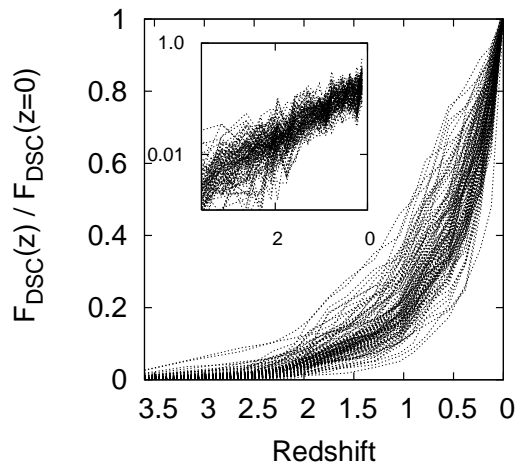,height=8.cm,angle=-90}
\caption{ Fraction of star particles found in the DSC at redshift
  $z=0$ which are already in the diffuse component at redshift $z$,
  for all clusters in our set. The inset show the same curves in
  logarithmic coordinates.  }
\label{f:summstrip}
\end{figure}

As already discussed, the star formation model used in our simulations
implies that stars can only form inside existing gravitational
potential wells, so that all DSC stars must have become unbound from
their parent galaxies sometime after their formation. In
Figure~\ref{f:summstrip}, we plot the fraction of star particles in
the DSC at $z=0$ which are already in the DSC at redshift $z$. The
bulk of the DSC is created after $z \approx 1$, when on average only
$\approx 30$ per cent of $z=0$ DSC star particles already reside 
outside their parent galaxies, with significant cluster-to-cluster
variations. However, from the inset of Fig.~\ref{f:summstrip} we note
that the production of the DSC follows a power-law, thus implying that
it is a cumulative process which, on average, does not have a
preferred time scale. \cite{Fabio} found a similar result based on the
analysis of their high--resolution simulation of a single cluster,
with a continuous growth of the DSC fraction and no preferred epoch of
formation.

No statistically significant correlation is found between the fraction
of DSC at $z=0$ and a number of possible tracers of the dynamical
history of the cluster, such as the concentration of the NFW profile,
the number of (DM--halo) major mergers, or the epoch of the last major
merger.  This suggests that the process of formation of the DSC is
more related to the local dynamics of the interactions between
galaxies and the group/cluster environment, rather than to the global
dynamical history of the cluster.

\subsection {DSC and the history of galaxies}

We now proceed to establish which galaxies are the main contributors to
the formation of the DSC. For each DSC star particle at $z=0$, we look
for a $G_j$ galaxy at $z_i$ to which it last belonged. When this
galaxy is found, we check whether the galaxy $G_j$ is associated with
the family tree of a galaxy $G_k$ at $z=0$. If so, then the DSC star
particle is associated with the ``family tree'' of the galaxy $G_k$.
If the $G_j$ galaxy at $z_j$ is not associated with the family tree of
any $G_k$ galaxies at $z=0$, but its family tree ends at $z_{j+m}$, then
the DSC star particle is associated with a dissolved galaxy. If no
bound structure is found, then the particle is associated with a {\sl
volatile} structure, and it is not considered in the subsequent
analysis.

As a next step, we compute what fraction of the DSC particles comes
from the family trees of galaxies at $z=0$, as a function of the
binned galaxy mass at $z=0$, $M_{\rm gal}(z=0)$. Then the DSC mass
$M^*_{\rm DSC}(M_{\rm gal})$ obtained for each $M_{\rm gal}(z=0)$ bin
is normalised by the total stellar mass of the respective cluster.
The total fraction $F_{\rm DSC}$ for each cluster is finally given by
the sum over all contributions from all galaxy masses at $z=0$.

\begin{figure*}
\centerline{\hfill\psfig{file=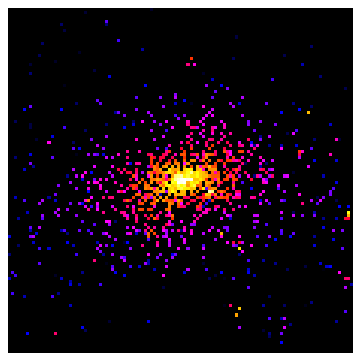,width=0.31\linewidth}
\psfig{file=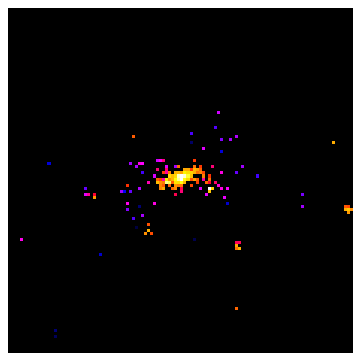,width=0.31\linewidth}
\psfig{file=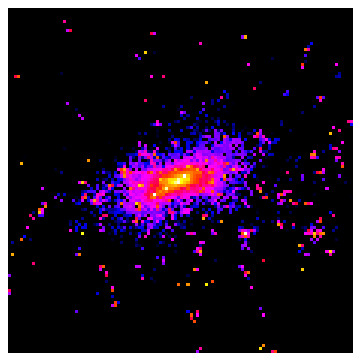,width=0.31\linewidth}\hfill}
\centerline{\hfill\psfig{file=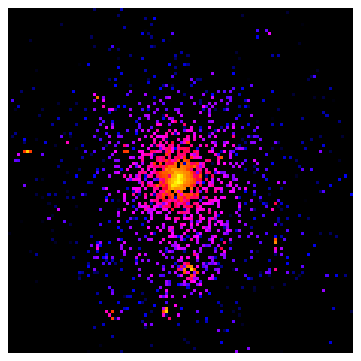,width=0.31\linewidth}
\psfig{file=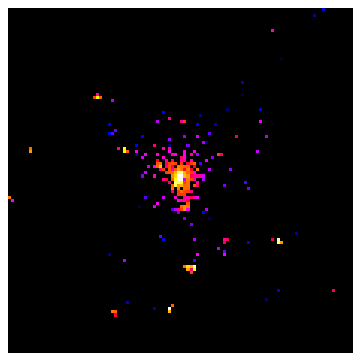,width=0.31\linewidth}
\psfig{file=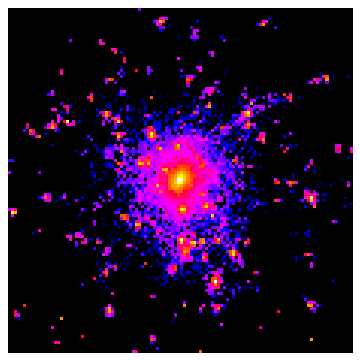,width=0.31\linewidth}\hfill}
\centerline{\hfill\psfig{file=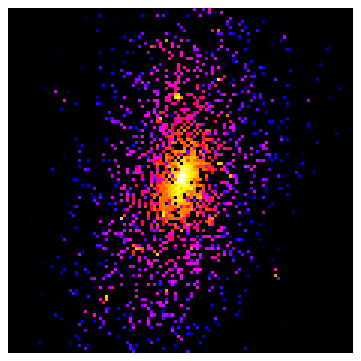,width=0.31\linewidth}
\psfig{file=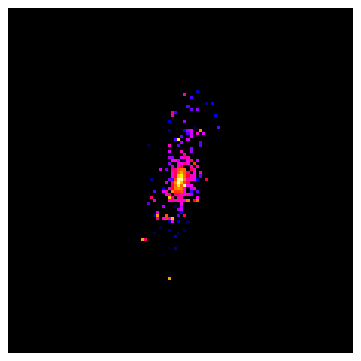,width=0.31\linewidth}
\psfig{file=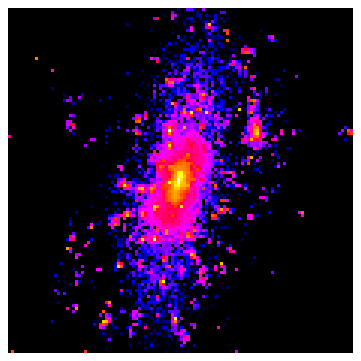,width=0.31\linewidth}\hfill}
\centerline{\hfill\psfig{file=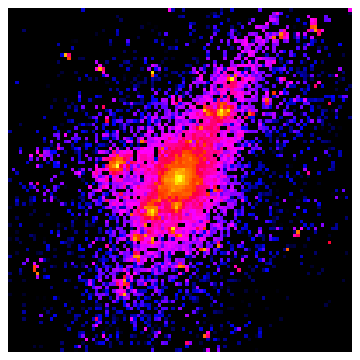,width=0.31\linewidth}
\psfig{file=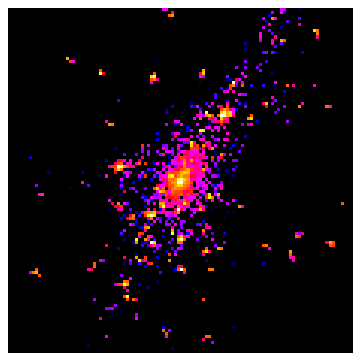,width=0.31\linewidth}
\psfig{file=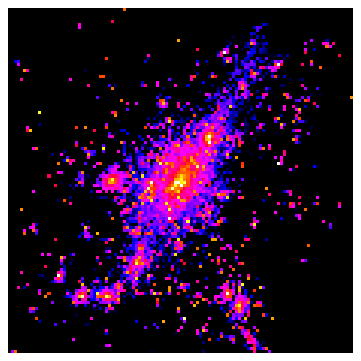,width=0.31\linewidth}\hfill}
\caption{The distribution of the dark matter (left panels) and of the
  stars (canter panels) for clusters (A), (B), (C), (D) from the
  cosmological simulation, and the distribution of stars in the
  high-resolution re-simulations of the same four clusters (right
  panels), all at redshift $z=0$. The frames are $3 h^{-1}$ Mpc on a
  side in the first three rows and $6 h^{-1}$ Mpc in the last row,
  corresponding to $\approx 2 R_{\rm vir}$ for the four clusters (see
  Table \ref{clustab}).  They show density maps generated with the
  SMOOTH algorithm, applied separately to the DM and star particle
  distributions. Colour scale is logarithmic and different for DM and
  stars: from $10^{-0.5}$ to $10^5$ times critical density and from
  $10$ to $10^6$ times critical density for stars and DM
  particles, respectively.  }
\label{4clus}
\end{figure*}

\subsection {Standard resolution simulation - 4 exemplary clusters }

We discuss the results of this analysis for the four clusters shown in
Figure~\ref{4clus}. The figure shows the density distribution of DM
and star particles in the four clusters, and also the distribution of
star particles in the high-resolution re-simulation of these clusters.
The main characteristics of these clusters are given in
Table~\ref{clustab}. The galaxies identified by SKID correspond to the
densest regions plotted in yellow in this Figure.

These four clusters cover a wide range of masses (see
Table~\ref{clustab}), and the two intermediate mass clusters
B and C have very different dynamical histories: cluster B
experienced a major merger at $z \approx 1$, while cluster C is
undergoing a merger event at $z=0$  which began at $z \approx 0.2$.

\begin{figure*}
\vspace*{8.cm}
\psfig{file=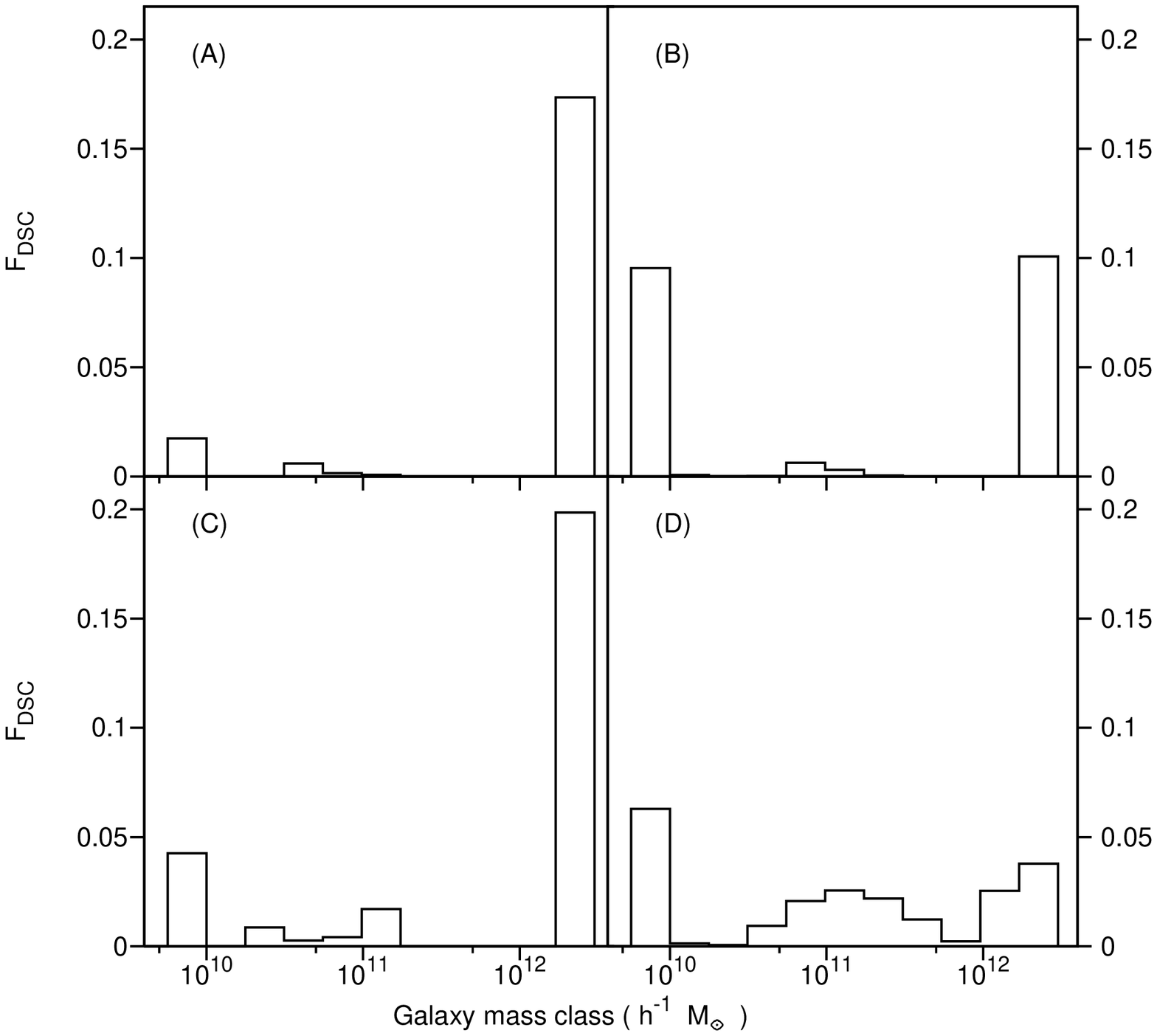,height=6.cm,angle=-90}
\caption{Histograms of the fraction of DSC star particles identified at $z=0$,
  associated with the family trees of $z=0$ galaxies of different masses:
  $F_{\rm DSC}(M_{\rm gal})=M^*_{\rm DSC}(M_{\rm gal})/M^*_{\rm tot}$. Results
  are reported for the four clusters shown in Figure \ref{4clus} (see also
  Table \ref{clustab}). We use 10 galaxy mass bins, logarithmically spaced,
  from $M_{\rm min}=1.1 \times 10^{10} h^{-1} M_\odot$ to $M_{\rm
  max}=3.1 \times 10^{12} h^{-1} M_\odot$. The leftmost column in each panel
  gives the contribution from dissolved galaxies, regardless of their mass.
  It is the mass range only for sake of clarity.  For these 4 clusters, the
  only family tree contributing to the rightmost column is that associated
  with the BCG.  }
\label{histosingleLR}
\end{figure*}

In Figure~\ref{histosingleLR}, the histograms show the mass fractions
of the DSC associated with the family trees of $M_{\rm gal}(z=0)$
galaxy for these four clusters.  
From Fig.~\ref{histosingleLR}, we can draw the following picture for the
origin of the DSC:
\begin{itemize}
\item The bulk of the DSC comes from the formation history of the most
  massive galaxy, except in the most massive cluster D;
\item Dissolved galaxies give a significant contribution in two
 out of the four clusters (clusters B,D);
\item All other galaxy family trees provide either a small (clusters
 A--C) or modest (cluster D) contribution to the DSC.
\end{itemize}

\begin{table*}
\caption{Virial masses, viral radii, and DSC fractions for the four clusters
A--D shown in Figure~\ref{4clus}. The fraction shown in the column 4, $F_{\rm
DSC}^{\rm all}$, includes the contribution from low-density galaxies. The
$F_{\rm DSC}$ value in column 5 is obtained omitting the particles unbound
from low-density galaxies.  For completeness we report the fraction $F_{\rm
DSC}^{\rm vol}$ of discarded particles from low-density or volatile structures
in column 6; the fraction $F_{\rm DSC}^{\rm dis}$ of star particles from
dissolved galaxies in column 7, which corresponds to the leftmost columns in
the histograms of Figs.~\ref{histosingleLR} and \ref{histoallLR}; the fraction
$F_{\rm DSC}^{\rm ng}$ of star particles that never belonged to any galaxy in
column 8. The last row of the table reports the average DSC fractions for the
whole sample of 117 clusters.  }
\label{clustab}
\begin{tabular}{c c c c c c c c}
\hline\hline Label & $M_{\rm vir}$ ($10^{14}h^{-1} M_\odot$) & $R_{\rm
 vir}$ ($h^{-1}$ kpc) & $F_{\rm DSC}^{\rm all}$ & $F_{\rm DSC}$ &
 $F_{\rm DSC}^{\rm vol}$ & $F_{\rm DSC}^{\rm dis}$ & $F_{\rm DSC}^{\rm
 ng}$\\ 
\hline 
A & 1.6 & 1200 & 0.33 & 0.20 & 0.11 & 0.02 & 0.02 \\ 
B & 2.5 & 1290 & 0.36 & 0.21 & 0.12 & 0.10 & 0.05 \\ 
C & 2.9 & 1350 & 0.45 & 0.27 & 0.14 & 0.04 & 0.04 \\
D & 13.0 & 2250 & 0.45 & 0.22 & 0.23 & 0.06 & 0.00 \\
Ave & -- & -- & 0.34 & 0.18 & 0.09 & 0.04 & 0.07 \\
\hline
\end{tabular}
\end{table*}

\begin{figure*}
\vspace*{8.cm}
\psfig{file=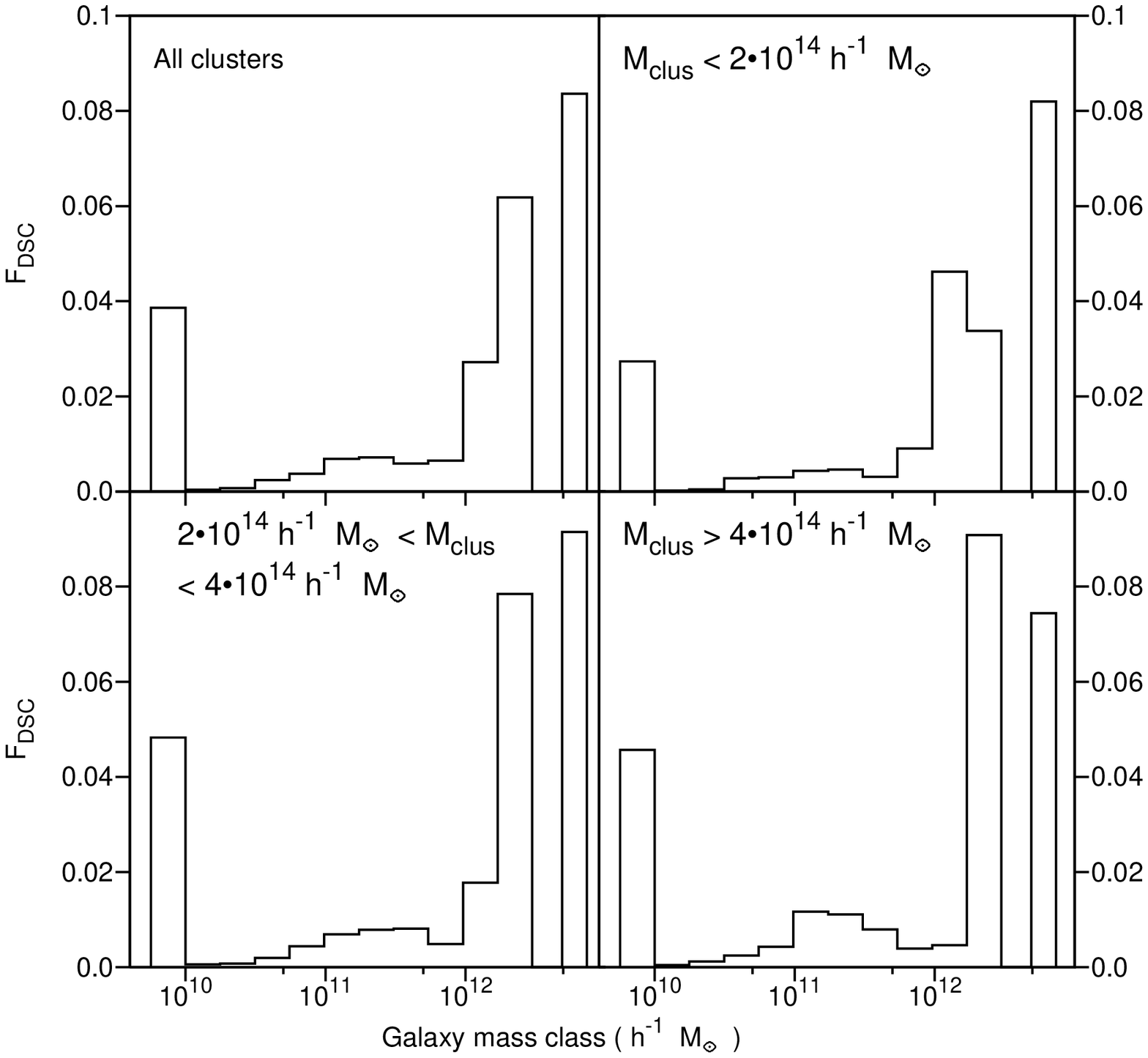,height=6.cm,angle=-90}
\caption{ Histograms of the relative contribution to the DSC from the
  formation history of galaxies belonging to 10 $M_{\rm gal}(z=0)$
  mass bins, for the entire set of clusters in the simulation, and as
  a function of cluster mass.  Upper left panel: average over the
  whole 117 cluster set. Upper right panel: average over the 71 least
  massive clusters. Lower right panel: average over the 11 most
  massive clusters.  Lower left panel: average over the 35
  intermediate--mass clusters.  We use 10 $M_{\rm gal}(z=0)$ mass
  bins, logarithmically spaced from $M_{\rm min}=1.1 \times 10^{10}
  h^{-1} M_\odot$ to $M_{\rm max}=3.1 \times 10^{12} h^{-1} M_\odot$.
  The leftmost column in each panel represents the contribution from
  dissolved galaxies, regardless of their mass. The rightmost column
  in each panel shows the contribution from the history of the single
  most massive galaxy of each cluster, regardless of its actual mass.}
\label{histoallLR}
\end{figure*}

In the case of the most massive D cluster, a significant contribution
to the DSC comes from intermediate--mass galaxies.  \cite{Fabio} also
found in their simulation of one cluster with mass similar to our
cluster D, that galaxies of all masses contributed to the production
of the DSC. Our results suggest that when the cluster statistics is
enlarged, such cases are rare; in our set this is the case in 3
clusters out of the 11 most massive ones from the whole set of 117
clusters.

Furthermore, Figure~\ref{4clus} shows that cluster D is still
dynamically young, with a number of massive substructures both in the
DM and in the star particle distribution.  This is probably the main
reason why the DSC formation in this cluster is not dominated by the
most massive galaxy: the sub--clumps contain galaxies of various
masses which experienced several mergers in their history, producing a
significant amount of DSC. This is also confirmed by the analysis of
the family trees of the galaxies belonging to this cluster: 12 of the
85 identified galaxies had more than one merger in their history,
while usually only one or two galaxies in each cluster are found to
have a complex family tree.

In the other 3 clusters, the largest fraction of the DSC star
particles is associated with the formation history of the cluster's
most massive galaxy. Cluster B also shows a large contribution coming
from dissolved galaxies: perhaps this suggests that the tidal field
associated in this cluster was more efficient in disrupting galaxies
rather than stripping some of their stars.

Our analysis so far does not exclude that some fraction of the DSC at
$z=0$ is produced in subclusters or groups, such as suggested by
\cite{Rudick}. 
In fact, the analysis of cluster D suggests that
this does happen. If these sub-clusters or groups migrate to the
centre of the cluster and finally merge, our procedure would associate
the DSC particles unbound from these structures with the family tree
of the cD at $z=0$.

However, if tidal stripping of the least--bound stars in all galaxies
were the main mechanism for the production of the DSC, we would expect
a more similar fraction of DSC star particles from all galaxy masses.

\subsection {Standard resolution simulation - statistics for 117 clusters }

We now turn to the statistical results for the whole set of 117
clusters.  In Figure~\ref{histoallLR} we show the contributions from
the same galaxy mass bins as before, but averaged over all clusters
and over different cluster mass ranges.  To obtain our average values,
we sum the mass of diffuse star particles in all clusters in the
appropriate galaxy mass bin and normalise it to the total stellar mass
of all clusters. This procedure creates a ``stacked-averaged''
cluster. The average value of the diffuse light fraction is $<F_{\rm
DSC}>=16 \%$ of the total stellar mass.

In the upper left panel of Fig.~\ref{histoallLR}, showing the fractional
contributions from galaxy mass bins averaged over the whole
cluster set, the rightmost column represents the
contribution from the clusters' BCGs only.  The value of the mass for
this class is arbitrary; this bin has been plotted separately since
the masses of the BCGs increase with cluster mass and, therefore, BCGs
in different clusters can belong to different mass bins. This effect is
clearly visible in the upper right panel of Fig.~\ref{histoallLR}, which
refers to the less massive clusters in our set. For these clusters,
the BCGs fall into two mass bins, with the majority of them falling
in the second most massive bin.

The other three panels of Fig.~\ref{histoallLR} show the same relative
contributions when the average is performed over (i) the 11 most
massive clusters (lower right panel, $M_{\rm vir} > 4 \times 10^{14}
h^{-1} M_\odot $ with $<F_{\rm DSC}>=19 \%$), (ii) the 35 clusters
having intermediate mass (lower left panel, $2 \times 10^{14} < M_{\rm
vir} < 10^{14} h^{-1} M_\odot$ with $<F_{\rm DSC}>=18 \%$), and (iii)
the 71 least massive clusters (upper right panel, $M_{\rm vir}< 2
\times 10^{14} h^{-1} M_\odot $ with $<F_{\rm DSC}>=13 \%$). These
average values show a weak trend with cluster mass in the production
of the DSC.  As a further test, we have divided clusters according the
amount of the DSC fraction itself. This analysis also confirms that
the dominant contribution to the DSC comes from the BCG family tree,
independent of $F_{\rm DSC}$ as expected, given the weak relation
between $F_{\rm DSC}$ and cluster mass. We also find that the
contribution from dissolved galaxies is slightly higher for clusters
with $<F_{\rm DSC}> $ greater than $25 \%$.

The results shown in Fig.~\ref{histoallLR} are consistent with the
previous analysis of the four clusters: the bulk of the DSC is
associated with the galaxies in the family tree of the most massive
galaxy of each cluster. Galaxies in the family trees of smaller $z=0$
mass bins contribute only few tenths of the fraction from the BCG
family tree.

Dissolved galaxies also contribute significantly to the DSC, but it is
possible that their estimated contributions are affected by some
numerical effects.  In fact, if the analysis is restricted to galaxies
whose density is within $1 \sigma$ of the observed galaxy density
estimate (Section~\ref{skidid}), then the contribution to the $F_{\rm
DSC}$ from the BCG rises to $\approx 76\%$, while that from dissolved
galaxies drops to $\approx 8\%$. As expected, most ``dissolved
galaxies'' are those with low densities, which indicates that their
contribution to the DSC may be affected by the limits in numerical
resolution. This needs further work to be properly understood.

\subsection {Merging and stripping in galaxy family trees}
\label{DSCmerging}

\begin{figure}
\psfig{file=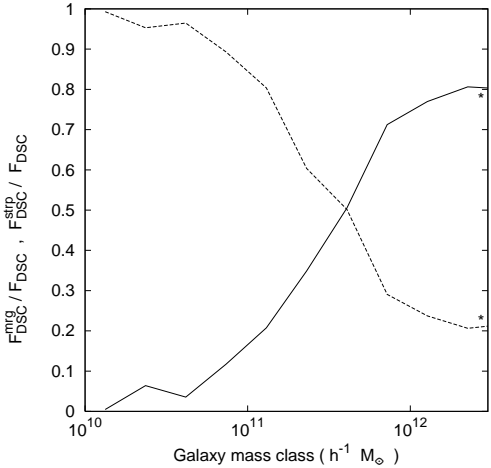,width=9.cm,angle=-90}
\caption{ The fraction of DSC stars arising in the ``merger'' part,
  $F_{\rm DSC}^{\rm mrg}/F_{\rm DSC}$, and in the ``stripping''
  part, $F_{\rm DSC}^{\rm strp}/F_{\rm DSC}$, of the galaxy family
  trees, as a function of their $z=0$ galaxy mass class (solid and
  dashed lines, respectively).  The asterisks mark the values for the
  contribution from the BCG family tree only. See text for details.}
\label{f.mrgstr}
\end{figure}

As shown in Fig.~\ref{famtreelow}, the BCG is the galaxy that
experiences the largest number of luminous mergers during its
assembly, and it often has the most complex family tree in a cluster.
Our result that a large fraction of DSC is associated with the BCG
supports a scenario where mergers release stars from their parent
galaxies to the intracluster space. 

To investigate this further, we estimate the fraction of DSC particles
associated with the {\sl merger} part of the family trees, and with the {\sl
stripping} part of the family trees, as follows.  We define a DSC star
particle to arise from a {\sl merger} at redshift $z_j$ if the galaxy it was
last bound to has more than one progenitor at
$z_{j-1}$\footnote{It may happen that the final phase of the merger is not
detected by our galaxy identification procedure, because the two merging
structures are very close to each other and have small relative velocities. In
this case, the SKID algorithm generally merges the two objects into a single
one, even if the merger is not yet completed. For this reason, we assign a DSC
star particle to the merger part of its parent galaxy family tree even if it
becomes unbound two family tree levels after the merger, at $z_{j+1}$
(i.e. from the ``offspring of the offspring'' of a merger).}.  The DSC
particles coming from the progenitors at $z_{j-1}$ are also defined as arising
from a merger. We take a DSC star particle to be unbound through {\sl
stripping} if the galaxy it was last bound to at redshift $z_j$ has only
one progenitor at $z_{j-1}$.  The different parts of the family trees are
indicated in Fig.~\ref{famtreelow}, where the squares represent the {\sl
merger} part of the tree, triangles the {\sl stripping} part, while circles
mark the part of the tree where no stars are released to the DSC.

In Figure~\ref{f.mrgstr}, we show the fraction of the DSC star
particles which arise from the {\sl merger} part of the family tree,
as a function of the final galaxy mass.  For high--mass galaxies,
most of the DSC originates from the {\sl merger} part of their family
trees. Low mass galaxies, on the other hand, lose stars only
via {\sl stripping}. After each {\sl merger} between massive galaxy
progenitors, up to $30\%$ of the stellar mass in the galaxies involved
has become unbound. This large fraction perhaps indicates that many of
these mergers take place either in strong tidal fields generated by
the mass distribution on larger scales, or just before the merger
remnants fall into their respective cluster.

Combined with the result that most of the DSC star particles are
associated with the family trees of the most massive galaxies, the
fact that most of the DSC is released during {\sl merger} events
implies that the bulk of the DSC originates in the merger assembly of
the most massive galaxies in a cluster.  The more standard picture for
the formation of the DSC, in which all galaxies lose their outer stars
while orbiting in a nearly constant cluster gravitational potential,
is not confirmed by current cosmological hydrodynamic simulations. It
appears that strong gravitational processes, linked to the formation
of the most massive galaxies in the cluster and to mergers between
luminous objects, are the main cause for the creation of the DSC.

A further mechanism possibly at work is the complete disruption of
galaxies, which also takes place preferentially in the cluster central
regions. In our cosmological simulation this formation mechanism for
the DSC is likely to be enhanced by numerical effects, which tend to
produce under-dense galaxies. We address this issue below when we
discuss our high-resolution simulations of the clusters in
Fig.~\ref{4clus}.

\subsection {cD Halo {\it vs} Intracluster Light}
\label{cdvsicl}

\begin{figure}
\vspace*{8.cm}
\psfig{file=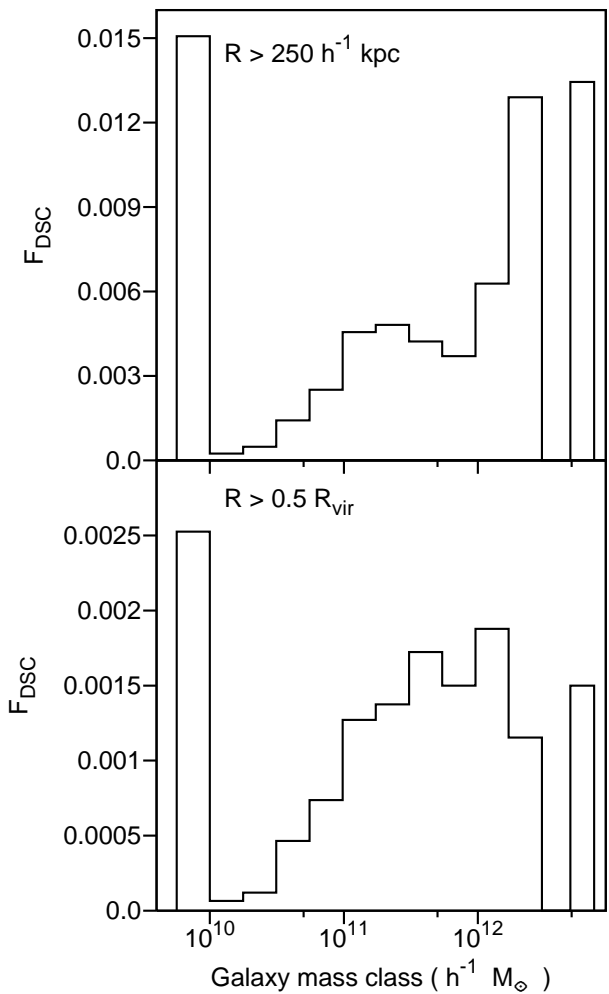,height=6.cm,angle=-90}
\caption{ The same as in Figure \ref{histoallLR}, averaging over
all clusters, but excluding the DSC star particles inside $R=250
h^{-1}$ kpc (upper panel) and inside $0.5 R_{\rm vir}$ of each cluster
(lower panel).  }
\label{nohalo}
\end{figure}

So far, we have made no attempt to distinguish between a component of DSC
associated with the unbound halos of the central cD galaxies, and a more
cluster-wide DSC.  Our definition of the DSC includes stars in the cluster
central regions and a part of these may well be in the form of cD halos (see
the Appendix).  Independent of how well a distinction between these two
components can be made, one might expect that the fraction of DSC stars that
comes from the merging tree of the BCG would be most concentrated towards the
cluster centre.  To shed some light on this question, we show in
Figure~\ref{nohalo} the same analysis for the average over all galaxy clusters
in the simulation as in Fig.~\ref{histoallLR}, but now excluding all DSC
particles residing in the central $250 h^{-1}$ kpc around their cluster
centres.  The remaining total DSC fraction drops to about $6 \%$, less than
half of the total, reflecting the steep radial profile of the DSC \citep{M04,
Zibetti}.  For $R>250 h^{-1}$kpc, the family trees of the most massive
galaxies still provide the largest contribution to the DSC (per mass bin), but
the cumulative contributions from family trees of less massive galaxies now
dominates the BCG component by a factor $\sim 2$. At the same time, the
relative contribution from dissolved galaxies increases. The lower panel of
Fig.~\ref{nohalo} shows the same analysis but now excluding DSC particles
within $0.5 R_{\rm vir}$. In this case, the $<F_{\rm DSC}>$ drops to $\sim
1\%$, with the fraction from the BCG family tree now similar to that from
other galaxies.

Since the BCG halo is likely to be less extended than $250 h^{-1}$ kpc, our
interpretation of the results in Fig.~\ref{nohalo} is that the {\sl merger}
part of the most massive galaxy family tree in each cluster contributes
substantially to the DSC also outside the cD halo. However, at radii $ 250
h^{-1} \mbox{kpc} < R < 0.5 R_{\rm vir}$, the cumulative contribution from the
family trees of other massive galaxies dominates the DSC.  Presumably, these
are the most massive galaxies within subgroups, which fell into the cluster
and brought in their own DSC, but which have not yet had time to merge with
the BCG. This interpretation is consistent with the simulation results of
\cite{Fabio} and \cite{Rudick}. Only in the outskirts of clusters, at $R> 0.5
R_{\rm vir}$, we find that the DSC particles come preferentially from the {\sl
stripping} part of family trees from all galaxy mass bins.

The relevance of {\sl merger} events for the formation of the DSC may
explain why diffuse light is more centrally concentrated than
galaxies, in both observations \citep{Zibetti,Magda02} and in
simulations (M04, \citealp{Fabio}, \citealp{SommerLarsen}).  Stars from
accreted satellite galaxies form extended luminous halos around
massive galaxies \citep{Abadi06}, and if these massive galaxies end up
concentrated to the cluster centre, their diffuse outer envelopes
would preferentially contribute to the DSC in the cluster centre.

Our results on the origin of the DSC are also consistent with the
predictions by \cite{Elena} that simulated fossil galaxy groups
have a larger amount of intra--group stars than normal groups. Indeed,
if fossil groups are the dynamically most evolved groups, then their
galaxies had more time to interact and build up the central elliptical
galaxy \citep[see e.g][]{Elena}. The number of galaxy--galaxy mergers
in groups appears to be closely related to the amount of DSC liberated
\citep{SommerGroups}.  A direct comparison of our results with
\cite{Fabio} is difficult because of their re-normalisation of the
simulated galaxy luminosities in order to fit the observed luminosity
function, but these authors also concluded that luminous galaxies
provide a substantial contribution to the DSC.

\section {High resolution simulations and the effects of numerical
resolution on the formation of the DSC}
\label{reseffect}

To address the stability of our results against mass and
force resolution, we have re-simulated three clusters extracted from
the cosmological box (A,B,C from Table \ref{clustab}) with 45 times
better mass resolution and $\simeq 3.6$ times smaller gravitational
softening.\footnote{Cluster D has been re-simulated at only 10 times
better mass resolution and its analysis is not discussed here.}  A
detailed presentation of these re-simulations is given in Borgani
et al. (2006). Galaxies formed in these high-resolution simulated
clusters have densities similar to those shown in Fig.~\ref{galdens}
for masses larger than $\approx 10^{11} h^{-1} M_\odot$, and the
low--density tail seen in Fig.~\ref{galdens} is shifted towards
lower masses, in accordance with the better resolution.

\begin{figure*}
\vspace*{8.cm}
\psfig{file=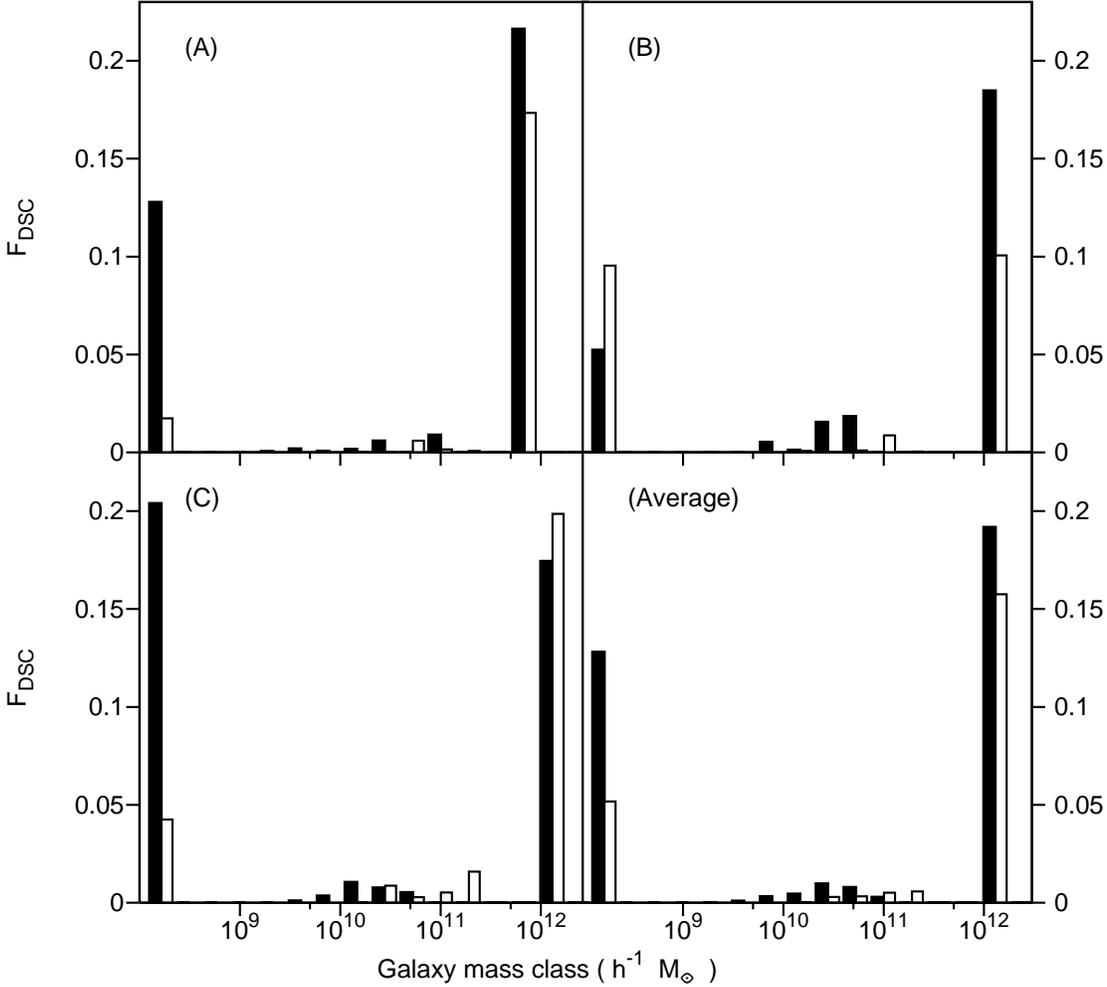,height=6.cm,angle=-90}
\caption{Histograms of the fraction of the $z=0$ DSC particles
associated with galaxy family trees in 15
 $M_{\rm
gal},(z=0)$ mass bins, for the three clusters re--simulated at
high--resolution (filled columns), and in the standard--resolution
(empty columns, cf.~Fig.~\ref{histosingleLR}).  The mass bins are
logarithmically spaced, from $M_{\rm min}=2 \times 10^{8} h^{-1}
M_\odot$ to $M_{\rm max}=3.1 \times 10^{12} h^{-1} M_\odot$.  The
leftmost columns show the contribution from dissolved galaxies,
regardless of their mass.  Upper-left, upper-right and lower-left
panels show the comparison for clusters (A), (B), (C), respectively;
the lower-right panel show the average for these three clusters.  }
\label{histrescomp}
\end{figure*}

We carry out our study of the DSC in these three clusters following
the procedure described previously, discarding DSC star particles
contributed from under-dense and volatile structures\footnote{For the
high--resolution clusters we use 24 different redshifts, starting from
$z=5$.}.  In Fig.~\ref{histrescomp}, we show the fractions of DSC star
particles identified at $z=0$ and originating from the history of
galaxies belonging to different mass bins. The full columns refer to
the analysis of the high--resolution simulations of the three
clusters, while the open columns show the results from
Fig.~\ref{histosingleLR} for the low-resolution simulations.

In the clusters simulated with high resolution, the results on the origin
of the DSC are consistent with what we found for the standard
resolution.  The DSC builds up in parallel with the formation of the
most massive galaxies in the cluster. The amount of DSC star particles
produced during the history of all other galaxies is still negligible
when compared with the contribution from the most massive cluster
members. With the increase in resolution, more DSC stars now come from
dissolved galaxies in clusters A and C, and an increasing number comes
from the family tree of the cD in clusters A and B.

Another question is whether the results on the DSC are affected by the
efficiency of the kinetic feedback from SNe. To address this point, we
re-simulate cluster A at the resolution of the cosmological box, with
(i) the same feedback efficiency and (ii) the speed of the galactic
ejecta set to zero.  We find very similar results in the strong
and weak feedback cases: no significant contribution from intermediate
mass galaxies, 7.5\% and 8.7\% of the DSC coming from dissolved
galaxies, and 89.7\% and 87.1\% of the DSC originating from the
history of the BCG, respectively.

However, the overall fraction of intracluster stars in these three
clusters changes between our low-resolution and high-resolution
simulations.  Once the star particles from volatile and under-dense
galaxies are discarded, the fraction of DSC within the virial radius
of clusters A, B and C is $F_{\rm DSC}=0.37, 0.28, 0.41$ in the high
resolution simulations, compared with $F_{\rm DSC} = 0.20, 0.21, 0.27$
in the standard resolution case. The overall increase of the DSC
fraction at high-resolution is mostly, but not only, related to
an increase in the fraction of DSC from dissolved lower-mass galaxies
(see Fig.~\ref{histrescomp}).

This result is not in contradiction with \cite{SommerGroups} finding
that the DSC fraction in his simulations remains constant when the
resolution is increased. This is because in the \cite{SommerGroups}
simulations the numerical resolution is increased without adding the
corresponding higher frequencies in the initial power spectrum. Then
the number of low-mass galaxies and (small--object) mergers does not
increase significantly, and one expects an approximately constant DSC
fraction. This suggests that the increase in the DSC fraction in our
high-resolution simulations is related to the addition of small objects
through the added high-frequency part of the power spectrum.

We expect that the effect of numerical overmerging is reduced at
higher resolution \citep[e.g.,] [ and references therein for a full
discussion of this issue]{BorgNum}, so we must look for other effects
that could dominate the disruption of the smaller galaxies.  The
results from \cite{SommerGroups} also rule out a significant effect
from stronger tidal shocks during high-speed collisions with
low-impact parameters, when galaxies become denser at higher
numerical resolution. One possibility is that, when the resolution is
increased, the number of numerically resolved mergers increases. On
the basis of the results reported above, this could turn into an
increased efficiency in the production of the DSC. If so, a solution
to the problem could lie in a more realistic feedback mechanism.

Another issue related to the numerical resolution concerns the number
of small ($\la 10^{11} h^{-1} M_\odot$) galaxies identified in these
simulations. Recent determinations of the K--band luminosity function
for galaxies in clusters give a faint--end slope between $\alpha =
-0.84$ and $\alpha=-1.1$ \citep{LinMohr04}.  The faint end slope of
the stellar mass function in our cosmological simulation is flatter
than the observed luminosity function: we obtain $\alpha \approx
-0.7$, thus implying that we miss a number of small galaxies. A
shallower slope of the faint end of the luminosity function is a
general problem of numerical simulations like those presented here
\citep[e.g.][]{Fabio}.

\begin{figure}
\psfig{file=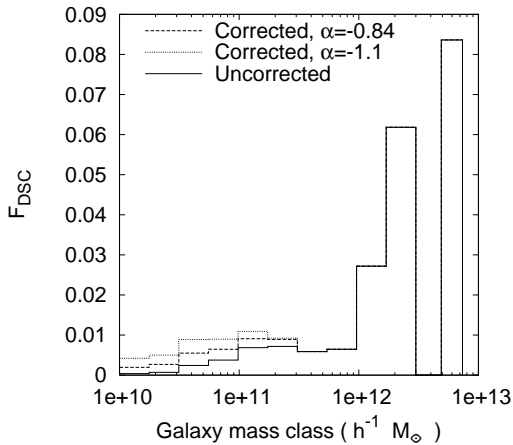,height=6.cm,angle=-90}
\caption{Corrections to $F_{\rm DSC}$ to account for the
low--mass end of the galaxy mass function.   The solid line is the
uncorrected $F_{\rm DSC}$; the dashed and the dotted lines give the
corrected value when the low--mass end of the galaxy cluster stellar
mass function has a slope $\alpha=-0.84$ and $\alpha=-1.1$,
respectively.}
\label{normfdsc}
\end{figure}

If the number of low--mass galaxies is underestimated in the simulations, then
their contribution to the DSC would be affected in the same way. To estimate
this effect, we computed how many galaxies we would expect in each mass bin of
Fig.~\ref{histoallLR}, if the faint end of the stellar mass function was given
by $n(M) = K \cdot (M/M_*)^\alpha$, with $\alpha=-0.84$,$-1.1$ and
$M_*=5\cdot10^{11} h^{-1} M_\odot$. The constant is fixed by requiring that
the number of galaxies for a mass $M=2\cdot 10^{11} h^{-1} M_\odot$ is the
same as in the simulation.  This method is similar to the re-normalisation of
the luminosity function in \citet{Fabio}.  We then assume that the missed
galaxies contribute the same relative fraction of their mass to the DSC as the
present-day galaxies of similar mass in the simulation, and multiply the
fraction $F_{\rm DSC}(M)$ by the ratio of $N(M)/N_s(M)$, where $N_s(M)$ is the
number of simulated galaxies found in the bin and $N(M)$ is the integral of
$n(M)$ in the same bin. This correction is applied to each mass bin up to
$2\cdot 10^{11} h^{-1} M_\odot$.  Fig.~\ref{normfdsc} shows the result of this
correction when it is applied to the average distribution of $F_{\rm DSC}$ for
the whole set of three clusters (lower right panel of Fig.~\ref{histrescomp}).
The DSC production is still dominated by the contribution coming from the BCGs
in the clusters.  The effect of such correction is to bring the contribution
of the mass bins corresponding to masses $M<10^{11} h^{-1} M_\odot$ to the
same levels of the others. Correction is stronger for the smaller mass
bins. Nevertheless, the contribution of these mass classes to the global
$F_{\rm DSC}$ remain small. Note that the increases in $F_{\rm DSC}(M)$ values
in the mass bin $\sim 2\cdot 10^{11} h^{-1} M_\odot$ is due to a few galaxies
with mass $1.5<M<2 \cdot 10^{11} h^{-1} M_\odot$, whose contribution to the
DSC has been corrected.  The contribution $F_{\rm DSC}(M)$ from galaxy having
masses smaller than $\approx 1.1 \cdot 10^{10} h^{-1} M_\odot$ in the three
re-simulated clusters, where they are resolved, is very small.

A further issue to be considered is that all galaxies are spheroidal at the
numerical resolution of these cosmological simulations; indeed, the
self--consistent formation of disk galaxies is still a challenge in
hydrodynamic $\Lambda$CDM simulations.  Are our conclusions on the origin of
the DSC likely to be affected by the absence of disk galaxies in our
cosmological simulation? Generally, disk galaxies are more vulnerable to tidal
forces, but the amount of matter lost in tidal tails is small, unless the
tidal field is very strong, whereas elliptical galaxies lose their outer stars
more easily. We do not expect that it would make a lot of difference for the
amount of DSC released in the merging processes leading to the formation of
the cluster BCG galaxies and most of the DSC, if a fraction of the
participating galaxies were disk galaxies. However, this needs to be checked
once simulations can reproduce disk galaxies.  Independent arguments based on
tidal stripping from disk galaxies in a semi--analytical model of galaxy
formation \citep{PG1,PG_MORGANA} suggest that at most $\sim 10$ per cent of the total
stellar mass of each cluster is contributed to the DSC by the `` quiet tidal
stripping'' mechanism, even for the most massive clusters where observations
points toward a larger amount of diffuse stars.

As already discussed in Section~\ref{skidid}, our force resolution is
not enough to resolve the inner structure of the simulated
galaxies. As a consequence, their internal density is likely to be
underestimated, so that these galaxies are more vulnerable to tidal
stripping and disruption than real galaxies. This numerical artifact
is not completely removed even in our high resolution re-simulations,
where the Plummer--equivalent softening is $\simeq 2 h^{-1}$ kpc. Even
when simulated galaxies with central densities lower than a chosen
threshold are removed, we still find that less massive galaxies are
less dense than massive galaxies, at variance with observational
results. This probably accounts for most of the contribution to the
DSC from dissolved galaxies.  If we had more realistic, denser small
galaxies in the simulation, this might decrease their contribution to
the DSC even more.

We can summarise the discussion of systematics effects in our analysis
related to numerical resolution as follows:
\begin{enumerate}
\item 
Our conclusion, that the formation of the DSC
is intimately connected with the build--up of the cluster's BCG, is
confirmed in higher numerical resolution simulations;
\item
this conclusion appears insensitive to the limitations of current
simulations in reproducing low-mass galaxies, disk galaxies, and
the faint-end luminosity function;
\item resolving galaxies with smaller masses in the high-resolution
simulations does not have a strong effect on the formation and
evolution of the DSC;
\item  the global value of $F_{\rm DSC}$ depends on resolution, and it
increases in simulated clusters with higher resolution. The value of
this fraction has not converged yet, in the range of numerical
resolution we examined.
\end{enumerate}

\section{How do stars become unbound? On the origin of the diffuse
stellar component}
\label{secunbound}

\begin{figure}
\centerline{\psfig{file=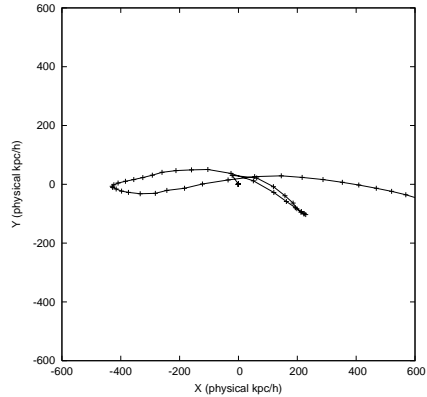,width=0.9\linewidth,angle=-90}}
\caption{Typical projected orbit of a galaxy ending in a major merger
with the cD galaxy; here in cluster A at
redshift 0.269. }
\label{figorbit}
\end{figure}

In cosmological hydrodynamic simulations, stars form in galaxies. The
DSC is built up from stars that are dissolved from their parent
galaxies. This is an ongoing process linked to the accretion of
substructure (see above and also \citealp{Fabio}); in fact, most
of the DSC originates at relatively recent redshifts.  Stars may be
added to the DSC through a number of dynamical processes, listed here
in the approximate time sequence expected during the infall of a
significant subcluster:

\begin{enumerate}
\item
Tidal stripping of the preexisting diffuse stellar population from the
in-falling subcluster or group: The DSC in the substructure, generated
by dynamical processes in the substructure, is added to the DSC of the
main cluster when both structures merge.
\item
Stripping from extended galaxy halos created in substructures: In
subclusters or galaxy groups, galaxy interactions occur with lower
relative velocities and are thus generally more damaging than in
interactions with the high relative velocities typical for galaxy
clusters.  Interactions or mergers within substructures may create
loosely bound stellar halos which are stripped from their parent
galaxies and the substructure when entering the cluster tidal field
\citep{Mihos04,Rudick}. This stripping may be delayed when
the merging in the subcluster happens before its accretion, or
immediate when the galaxy interactions occur already deep in the tidal
field of the main cluster.
\item 
Tidal shocking and stripping during merger with the cD galaxy: The
massive galaxies in the substructure generally interact with the
cluster centre and the cD galaxy on near-radial orbits (see
Figure~\ref{figorbit}). In a high-speed encounter of a massive galaxy
with the cD, stars from both galaxies may gain sufficient energy to be
(almost) tidally unbound from their parent galaxies. The tidally
shocked stars from the intruder are then subsequently unbound by the
ambient tidal field or further tidal shocks, remaining at similar
orbital energies as their mother galaxy had at the time. Those from
the cD galaxy become part of the cD envelope. The process may happen
several times as the intruder galaxy orbit decays by dynamical
friction.  This mechanism is related to the cannibalism scenario for
the growth of the cD galaxy, described in \cite{Ostr77, Merritt95}
 and others; however, here the dynamical
friction appears to be more effective, presumably due to the large
dark matter mass associated with the infalling substructure.
\item 
Tidal dissolution of low-density galaxies: These galaxies may enter
high-density regions of the cluster along their orbits, such as the
dark matter cluster centre, and dissolve completely if of sufficiently
low density. 
\item 
Tidal stripping in galaxy interactions: Stars may be torn out from
galaxies during tidal interactions along their orbits in the cluster,
and be dissolved from their parent galaxies by the cluster tidal
field. The participating galaxies survive as such. Galaxies of all
masses are affected. The last two processes together are often
described as harassment \citep{Moore96}.
\end{enumerate}

The statistical results of the previous sections allow us to put
some constraints on the relative importance of these various 
processes, and to identify further work needed to clarify the
origin of the diffuse stellar population in galaxy clusters. 
These results can be summarised as follows. 
\begin{itemize}
\item
Most of the DSC is liberated from galaxies in the merger tree of the
most massive, central galaxy in the cluster, i.e., simultaneously with
the build-up of this galaxy.
\item
If only the fraction of the DSC outside $250{\rm kpc}$ from the
cluster centre, i.e., outside the cluster dominant galaxy's halo, is
considered, the contribution from the BCG family tree is comparable to
that from other massive galaxies.  Only outside $\sim 0.5 R_{\rm vir}$
do galaxies of all masses contribute to the DSC.
\item
There is a further, sizable contribution to the DSC in the
simulations, from dissolved galaxies. However, the fraction of DSC
stars from dissolved galaxies depends directly on the simulations'
ability to faithfully represent the lower mass galaxies, and is seen
to vary strongly with the resolution of the simulation.  This
contribution to the DSC is thus currently uncertain; the prediction
from the simulations is likely to overestimate the contribution of
dissolved galaxies to the ICL in observed clusters.
\item
About 80\% of the DSC that comes from the merger tree of the cD
galaxy, is liberated shortly before, during, and shortly after major
mergers of massive galaxies. About 20\% is lost from these galaxies
during quieter periods between mergers.
\item In each significant merger, up to $30\%$ of the stellar
mass in the galaxies involved becomes unbound.
\item Most of the DSC is liberated at redshifts $z=0-1$.
\end{itemize}

These results imply that the main contribution to the DSC in our
simulations comes from either tidal shocking and stripping during
mergers with the cD galaxy in the final cluster (mechanism iii),
and/or from merging in earlier subunits whose merger remnants later
merge with the cD (i, ii).  The traditional tidal stripping process
(v) appears to contribute only a minor part of the DSC but may be the
dominant process for the small fraction of the DSC that ends up at
large cluster radii.

Further work is needed to see which of the channels (iii) or (i,ii) is
the dominant one, and whether in the latter the contribution of the
preexisting DSC in the accreted subclusters (i) dominates over that
from extended galaxy halos (ii).  The work of \cite{Fabio} and \cite{Rudick} 
shows that the contribution from infalling
groups is important but the division between the channels (i) and (ii)
is not clear. In this case, our results imply (a) that merging must
have been important in these groups, and (b) that the massive galaxies
in the infalling groups that carry most of the final DSC will mostly
have merged with the BCG by $z=0$, so that the tidal shocking and
stripping process (iii) will play some role as well.  The description
of \cite{Elena} and \cite{SommerGroups} of fossil groups
as groups that are older than other groups, in a more advanced
evolutionary stage with a dominant elliptical galaxy, and with a
larger fraction of DSC, suggests dense, evolved groups as promising
candidates for contributing significantly to the cluster DSC.  One
recent observational result that also fits well into this picture of
accreting groups that have already had or are having their own merger
events, is the observation of \cite{Aguerri2006} that the DSC
fraction in Hickson groups correlates with the elliptical galaxy
fraction in these groups.

Certainly it is clear from our results that the formation of the cD
galaxy, its envelope, and the DSC in galaxy clusters are closely
linked. Further analysis is required to determine whether these
components are dynamically distinct, and what kinematic signatures can
be used to distinguish between them in observations of cD clusters.

\section{Conclusions}
\label{concl}
In this paper, we have studied the origin of the diffuse stellar
component (DSC) in galaxy clusters extracted from a cosmological
hydrodynamical simulation.  We identified galaxies in 117 clusters
with the SKID algorithm, tracing each of them back in time at 17
different redshifts from $z=3.5$ to $z=0$. This allowed us to build
the family tree of all galaxies identified at $z=0$ in all clusters. We
find that all BCGs are characterised by complex family trees, which
resemble the merging trees of DM halos. At the resolution of our
simulation, only a small number of massive galaxies other than the
BCGs undergoes several mergers during their past history. The majority
of galaxies never have mergers, or only one at very early redshift.

Because of the star formation criteria employed in the simulation, all
stars found in the DSC at $z=0$ were born in galaxies and later
dissolved from them. We track each DSC star particle back to the last
redshift when it still belonged to a galaxy, and thus link it to the
dynamical history of this galaxy. We exclude all DSC star particles from
the analysis which arise from volatile and under-dense galaxies; the
latter being defined relative to the observational mass--radius
relation of early type galaxies by \cite{Shen}.
The main results of our analysis can be summarised as follows.
\begin{itemize}
\item The formation of the DSC has no preferred redshift and is a
  cumulative power--law process up to redshift $z=0$. We find that
  $\simeq 70$ per cent of the DSC is formed after redshift $z=1$.
\item We find a weak increase of the final amount of DSC stars with
  the mass of the cluster, but no significant correlation with the
  global dynamical history of the clusters.
\item For all but the 3 most massive clusters, DSC star particles come
  mainly from the family tree of the most massive (BCG) galaxy. I.e.,
  the formation of the DSC goes largely in parallel with the build-up
  of the BCG galaxy.
\item Most DSC star particles become unbound during merging phases along
  the formation history of the BGCs, independent of cluster mass.
\item Masking the inner 250 $h^{-1}$ kpc of each cluster, in order to
  exclude the cD halo from the analysis, does not qualitatively change
  the emerging picture.
\end{itemize}

From these results we conclude that the bulk of the DSC star particles
are unbound from the galaxy in which they formed by the tidal forces
acting before, during, and shortly after merging events during the
formation history of the BCGs and other massive cluster galaxies.
Only in the outskirts of clusters, $R > 0.5 R_{\rm vir}$, we find that
galaxies of many different masses provide comparable contributions to
the DSC , which is similar to a ``quiet stripping'' scenario, but the
actual mass in DSC stars in these regions is small.

The formation of the BCG in these simulations is related to many
mergers which begin early in the history of these galaxies and
continue all through $z=0$.  As discussed in the previous section, it
is reasonable to infer that the massive elliptical galaxies, which
merge with the BCGs, are contributed by infalling groups, which have
already generated their own DSC and/or loosely bound halos, as found
by \cite{Fabio} and \cite{Rudick}. Part of the cluster
DSC will also be generated by the tidal shocking and stripping during
the merger of these massive galaxies with the BCG itself; the relative
importance of these processes is yet to be established.

Since the fraction of diffuse light stars contributed by each
accreting group depends on the details of the dynamical history of the
group itself, such a mechanism for the generation of the DSC can hide
a direct link with the formation history of the clusters. This may be
the reason why we do not detect a clear correlation between the $z=0$
DSC fraction and the cluster formation history.

At the resolution of our (and other similar) simulations, it is not
yet possible to resolve the inner structure of low-mass galaxies. We
have taken this into account in our analysis by discarding all DSC
particles from galaxies with densities below a threshold set by
observations. In addition, there are well-known problems in
cosmological simulations with forming disk galaxies, and with
reproducing the galaxy luminosity function. These issues clearly
introduce some uncertainty in the discussion of the origin of the DSC
in hydrodynamic $\Lambda$CDM simulations.  We find that the global
amount of DSC in our simulations {\sl increases} with numerical
resolution and has not yet converged in the best simulations. Thus a
straightforward comparison of observed DSC fractions with numerical
simulations is not possible yet. On the other hand, massive galaxies
are well-resolved in our simulations, and we believe that our main new
result, that a major fraction of the DSC in galaxy clusters is
dissolved from massive galaxies in merging events, is a robust one.

\section*{Acknowledgements}

The simulations were carried out at the ``Centro Interuniversitario
del Nord-Est per il Calcolo Elettronico'' (CINECA, Bologna), with CPU
time assigned under INAF/CINECA and University-of-Trieste/CINECA
grants. This work has been partially supported by the PD-51 INFN
grant.  OG thanks the Swiss National Science Foundation for support
during the early stages of this work under grant 200020-101766.  We
acknowledge financial support by INAF projects of national interests
(PI: M. A.). KD acknowledges support by a Marie Curie fellowship of
the European Community program ``Human Potential'' under contract
number MCFI-2001-01221.  We acknowledge S. Bonometto, P. Monaco and
L. Tornatore for fruitful discussions and V. Springel for kindly
providing us the non--public part of the GADGET code.

\bibliographystyle{mn2e}
\bibliography{master}

\newpage

\section*{APPENDIX A: Identifying cluster galaxies with SKID} 
\label{AppA}

For the purposes of the present work, we need a dynamical and
automated way to identify galaxies in the simulations. Our galaxies
must be self---bound structures, locally over-dense, and we need an
operational procedure to unambiguously
decide whether a star particle at a given redshift belongs to an
object or not. For this reason, we follow the procedure adopted in M04
and use the publicly available SKID algorithm \citep{SK} and
apply it to the distribution of star and dark matter particles. 

\begin{figure*}
\centerline{\hfill\psfig{file=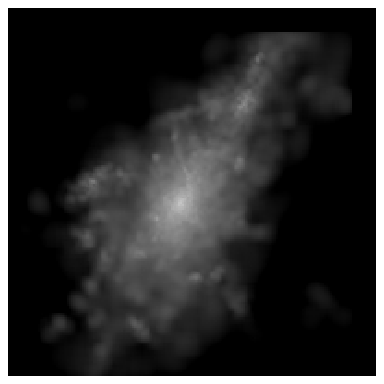,width=0.42\linewidth}
\psfig{file=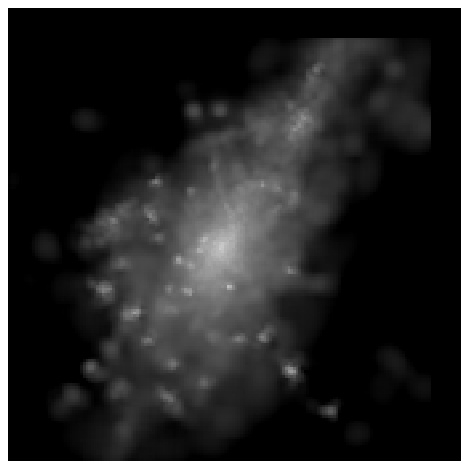,width=0.42\linewidth}\hfill}
\caption{ Surface density map of the DSC found in cluster D, when
three values of $N_{\rm sm}$ are used (left panel) and when only one
is used (right panel).  }\label{appmaps}
\end{figure*}

\begin{figure}
\vspace*{12.cm}
\psfig{file=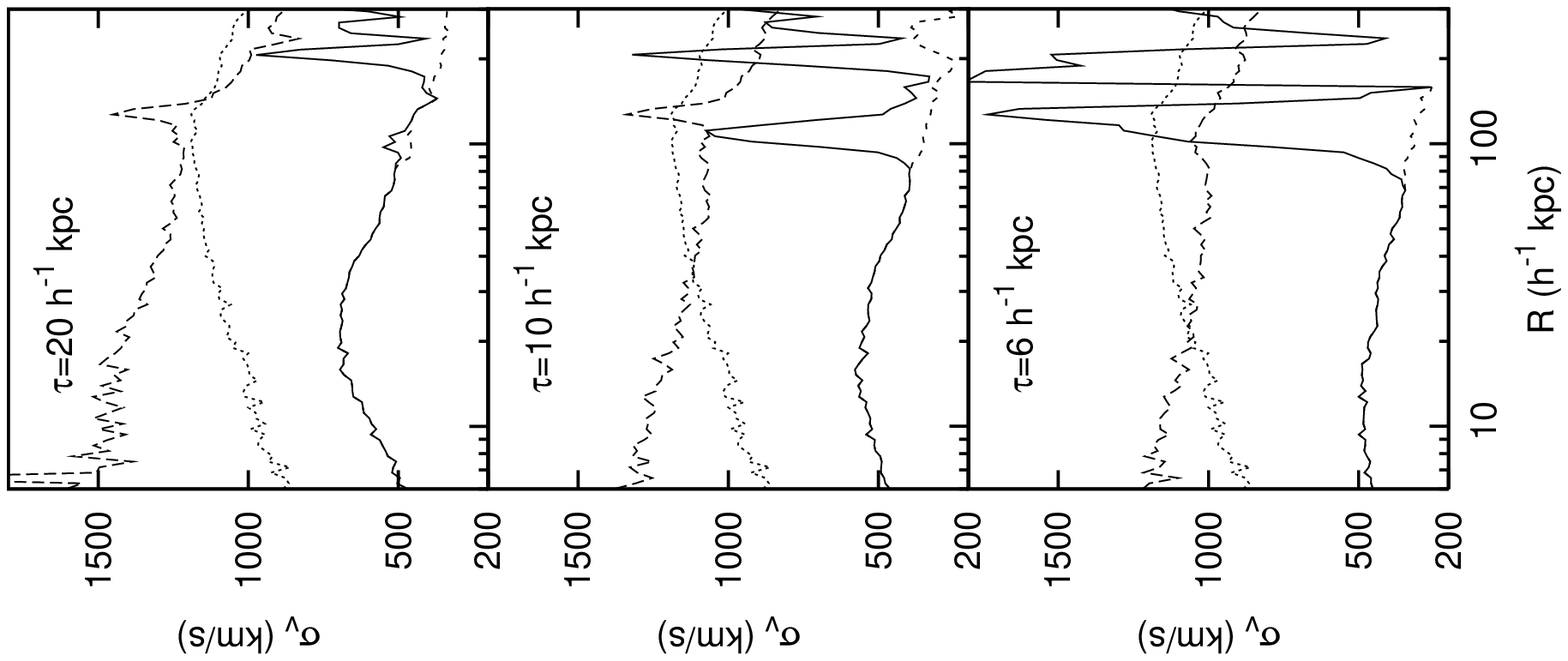,height=5.cm,angle=-90}
\caption{ 
Velocity dispersions in the central part of cluster A, simulated at our
higher resolution. Solid lines: stars belonging to SKID galaxies;
long-dashed lines: DSC stars; dotted lines: DM particles; short-dashed
lines: cD stars. Upper panel shows the results for the SKID analysis
performed with a value of $\tau=20 h^{-1}$ kpc, centre panel for
$\tau=10 h^{-1}$ kpc, lower panel for $\tau=6 h^{-1}$ kpc.
}
\label{sigmavel}
\end{figure}

The SKID algorithm works as follows:
\begin{itemize}
\item An overall density field is computed from the distribution of
  all available particle species, generally DM, gas and star
  particles.  The density is estimated with a SPH spline--kernel,
  using a given number $N_{\rm sm}$ of neighbour particles. In the
  following we only include DM and star particles.
\item The star particles are moved along the gradient of the density
  field in steps of $\tau/2$. When a particle begins to oscillate
  inside a sphere of radius $\tau/2$, it is stopped. In this way,
  $\tau$ can be interpreted as the typical size of the smallest
  resolved structure in the distribution of the star particles.
\item When all star particles have been moved, they are grouped using a
  friends-of-friends (FOF) algorithm applied to the moved particle
  positions. The linking length is again $\tau/2$.
\item The gravitational potential and binding energy of each group
  identified in this way is computed by accounting for all the
  particles inside a sphere centred on the centre of mass of the
  group and having radius $2\tau$ (for the moved star particles, their
  initial positions are used in the computation of the potential). The
  binding energies of individual particles are then used to remove
  from the group all the star particles which are recognised as
  unbound, in an iterative way: the centre of mass of the group and
  its potential are recomputed after a particle has been discarded.
\item Finally, we retain such a SKID--group of stars as a galaxy if it
  contains at least 32 particles after the removal of unbound
  stars. The exact value of this number threshold is unimportant, but
  the smaller the threshold is, the higher is the probability of
  identifying as ``galaxy'' a random set of neighbouring star
  particles. Using 32 particles correspond to a mass threshold of
  $M=1.1 \times 10^{10} h^{-1} M_\odot$ for the cosmological simulation. 
\end{itemize}

The resulting set of objects identified by SKID depends on the choice
of two parameters, namely $\tau$ and $N_{\rm sm}$. After many
experiments and resorting to visual inspection in a number of cases,
we find that a complete detection of bound stellar objects requires
the use of a set of different values of $N_{\rm sm}$. Using only one
value for $N_{\rm sm}$ results in ``missing'' some galaxies. We use
$N_{\rm sm}=16,32,64$, and define a {\sl galaxy} to be the set of star
particles which belong to a SKID group for any one of the above
$N_{\rm sm}$ values. If a star particle belongs to a SKID group for
one value of $N_{\rm sm}$ and to another group for a different $N_{\rm
sm}$, then the groups are joined and are considered as a single
galaxy. All star particles not linked to any galaxies are considered
to be part of the diffuse stellar component in the cluster. The left
panel of Fig. \ref{appmaps} shows the surface density map of the DSC,
as identified for our cluster D when all the three values of $N_{\rm
sm}$ are used. In the right panel, we show the same map obtained using
only $N_{\rm sm}=32$. In the latter, the bright spots correspond to
``missed'' galaxies.

$\tau$ roughly corresponds to the size of the
smallest resolved structure, and we adopt
$$
\tau \approx 3 \epsilon \eqno(A.1)
$$ 
which is the scale where the softened force becomes equal to the
Newtonian force. We have tested this choice by
visual inspection in a number of clusters, and by performing an
analysis of the velocity dispersions for the stars belonging to SKID
galaxies and to the DSC. Fig. \ref{sigmavel} shows the velocity
dispersions for various components, namely the stars in galaxies, the
stars in the DSC, the stars in the cD galaxy and the DM particles, for
our cluster A (re-simulated at our higher resolution) when the value of
$\tau$ is varied. The velocity dispersions are computed in spherical
shells centred on the cluster centre, defined as the position of the
DM particle having the minimum gravitational potential. For this
high-resolution simulation, our fiducial choice is $\tau=6 h^{-1}$
kpc. In the bottom panel of Fig. \ref{sigmavel}, the spikes in the
velocity distribution of the stars in galaxies (the solid curve) at
$R>100 h^{-1}$ kpc correspond to SKID objects; no prominent spikes or
bumps appear in the velocity dispersion profile for the DSC
(long-dashed line), meaning that no structures in velocity that might
correspond to ``missed'' galaxies are present in this component. Also,
the value of the velocity dispersions for DSC and DM particles (dotted
line) stay within $\approx 20$ \% from each other, as expected when
both component sample the same gravitational cluster potential.

When a larger value is used ($\tau=10 h^{-1}$ kpc), spikes begins to
appear in the DSC velocity dispersion curves, indicating that some
objects, or part of them, are missed by the algorithm. This is
especially clear for the spike at $R \approx 100 h^{-1}$ kpc, which is
present both in the velocity dispersion of stars in galaxies and in
the velocity dispersion of the DSC, and clearly indicates that a
fraction of stars in some galaxies have been mis-assigned to the
diffuse component. Also, the discrepancy between DSC and DM velocity
dispersions begins to grow, and the velocity dispersion of cD stars
gets unrealistically large, $\sigma_v> 500$ km/s. This happens because
some low--speed DSC stars begin to be assigned to the cD, whose
typical star particle velocities are even lower, thus increasing the
velocity dispersion of the cD star population. The situation gets
worse it the value of $\tau$ is increased to $20 h^{-1}$ kpc.

We have performed a similar analysis on some clusters taken from our
cosmological set, where the fiducial value from eq. (A.1) is $\tau=20
h^{-1}$ kpc. Again, increasing $\tau$ to larger values results in
having structures in the velocity space of the DSC, presumably due to
missed objects. We conclude that the scaling (A.1) gives good results
for the SKID galaxy identification, while keeping $\tau$ fixed to a
given value (e.g. $20 h^{-1}$ kpc) when the force resolution is varied
is not a good choice.

When analysing particle distributions at redshift $z>0$, we keep fixed
the value of $\tau$ in co-moving coordinates, thus allowing the
minimum physical size of our object to decrease with increasing
redshift. While this does not obey equation~(A1), $\tau$ never becomes
less than $\epsilon$. The effect is probably to slightly increase the
amount of ``volatile'' galaxies at higher redshifts.  Again, this
choice was tested by visual inspection and by analysing the velocity
dispersion distributions.

Also, we note that at high redshift ($z>1$) the distribution of gas
particles inside star--forming proto--clusters often contains adjacent
clumps of star particles. Applying SKID to such distribution results
in ``galaxies'' composed of two or more of such clumps, which instead
should be considered as separate galaxies. For this reason, we choose
to use only DM and star particles for the galaxy identification.

\label{lastpage}

\end{document}